\documentclass[12pt]{article}
\usepackage{amssymb}
\usepackage{graphics}
\usepackage{amsmath}
\numberwithin{equation}{section}
\begin{document}
{\bf
\title{What can we learn studying black holes?}}
\vspace{30mm}
\author{ Victor Berezin \\Institute for Nuclear Research of
Russian Academy of Sciences \\ berezin@ms2.inr.ac.ru }
\maketitle
\vspace{5mm}
\begin{abstract}
In this paper we try to answer the main question: what is
a quantum black hole?
\end{abstract}
\newpage
\section{Introduction}

What does urge a researcher to investigate quantum black
holes? Honestly, his own intrinsic interest. Besides, there
are other, more (or less) scientific reasons. It is commonly
believed that only small black holes can be considered as
quantum objects. Small, what does it mean? To estimate, we
should compare the size of the black hole with the
corresponding Compton length. The gravitational radius $r_g$
of the black hole of mass $m$ equals $r_g = \frac{2Gm}{c^2}$, where
$G$ is the Newtonian gravitational constant, and $c$ is the
speed of light. The Compton length of such particle is
$\lambda = \frac{\hbar}{mc}$ ($\hbar$ is the Planck's constant). 
If $r_g \simeq \lambda$, than the so called Planckian mass is 
$m_{pl}= \sqrt{\frac{\hbar c}{G}} \sim 10^{-5} gr$, the
Planckian length is $l_{pl} = \sqrt{\frac{\hbar G}{c^3}} \sim 10^{-33} cm$. 
In what follows we us the
units in which $\hbar = c=1$, so $m_{pl}=1/\sqrt{G}$, $l_{pl}=\sqrt{G}$. 
Black holes of so
small mass and sizes could be created from large metric
fluctuation in the very early Universe (primordial black
holes), or during vacuum phase transition.

The mankind ``knows'' about black holes  since 1784, when some priest
J. Michell recognized \cite{1} that stars becomes invisible if the correspond
escape velocity will exceed the speed of light. He called such stars
``dark stars'' Fifteen years later, in 1799, P.S. Laplace derived a relation
between the radius and the mass of these dark stars \cite{2}.

Let us consider a spherically symmetric gravitating 
body of mass $M$ and of radius 
$R_0$. Then the escape velocity $v_e$ can be determined from the balance of 
the kinetic and potential energies of the particle that starts to move away 
from this gravitating body. Using the non-relativistic mechanics and 
non-relativistic Newtonian theory of gravity we obtain, following Laplace 
($G$ is the Newton constant, and $m$ is the mass of the escaping particle) 
\begin{eqnarray}
\frac{m v_e^2}{2} = \frac{G m M}{R_0} ,\nonumber\\
\\
v_e^2 = \frac{2 G M}{R_0} .\nonumber
\end{eqnarray}
Equating the escape velocity to that of light $c$ we get for the maximal 
radius of the invisible body
\begin{equation}
R_0 = \frac{2 G M}{c^2} .\nonumber
\end{equation}

We are used to think of the black holes as of the extremely dense bodies. 
Indeed, the black hole with the mass of the Earth has the radius about 3 cm 
and the density much higher than the nuclear density. But due to the linear 
radius-mass relation the mean density $\rho$ is proportional to the inverse 
mass squared,
\begin{equation}
\rho = \frac{3 c^6}{32 \pi G^3 M^2} .\nonumber
\end{equation}
And for the huge black hole formed, say, of the cluster of galaxies, the 
density may be less than that of the atmospheric air. Thus the observer 
falling into such a black hole would not recognize that he never would go out 
of it (moreover, the free fall time in this case is about 10,000 years). 

The history of the ``real'' black holes started in 1916, when K.Schwarzschild
found the, now famous, solution to the Einstein equation for the spherically
symmetric vacuum space-time outside gravitating point source \cite{3}.
He derived  it in the form
\begin{eqnarray}
\label{schw}
ds^2 =  Fdt^2 - F^{-1}dr^2 - r^2 d\Omega^2 \\
\\
\nonumber
F = 1 - \frac{2Gm}{r}
\end{eqnarray}
Here $d\Omega^2$ is the line element of the unit sphere, $r$ - its radius,
and  $m$ is the total mass (energy) of the system, also known as Schwarzschild
mass. The line element (\ref{schw}) has a singularity at the so called
Schwarzschild radius  $r_g = 2 G m$, from where the light cannot escape to
the infinity. It is surprising that value of the Schwarzschild radius calculated
 using General Relativity coincides exactly with  non-relativistic (Newtonian)
radius of the dark stars.
 
In classical General Relativity black holes are very special
(and, therefore, very interesting) objects. First of all,
they are universal in the sense that they are described by
only few parameters: their mass, angular momentum and
charges. In the process of black hole formation, (i.e., in
the process of gravitational collapse) all higher momenta
and non conserved charges are radiating away. This
feature is formulated as a following conjecture: "black
holes has no hairs". Thus, a black hole formation results
generically in the loss of information about initial states
and previous history of collapsing matter. The boundary of
the black hole, the so-called event horizon, is the null
hyper-surface that acts as a one-way membrane. The matter can
fall inside but can not go outside. Because of this the area
of black hole horizon can not decrease. These two features,
the loss of information and nondecrease of the horizon
area, allowed J.Bekenstein \cite{3} to suggest the analogy
between the black hole physics and thermodynamics and
identify the area of the horizon with the entropy (up to
some factor). He did this for the simplest, spherically
symmetric neutral (Schwarzschild) black hole which
characterized by only one parameter, Schwarzschild mass.
Later the four laws of thermodynamics were derived for a
general black hole.

In thermodynamics the appearance of the entropy is
accompanied by the temperature. While the nature of the
black hole entropy was more or less clear, the notion of its
temperature remained mysterious until the revolutionary work
by S.Hawking \cite{5}. He showed that the black hole
temperature introduced by J.Bekenstein is the real
temperature in the sense that the black hole radiates, and
this radiation has a Planck's spectrum. The entropy
appeared
equal one fourth of the event horizon area divided by
Planckian length squared. Thus, even large (compared to the
Planckian mass and size) black holes exhibit quantum
features. It should be stressed that such quantum effect is
global, namely, it emerges as a result of nontrivial
boundary
conditions for the wave function of a quantum field theory
in curved space-times nontrivial causal structure (existence
of the event horizon(s)). 

Due to the process of Hawking's evaporation any (even
super-large) black hole becomes eventually small enough to
be considered as a (local) quantum object. At this stage the
combination of the global and local quantum effects may
result in unpredictable features. That why it is so exciting
to try to understand quantum black hole physics. 

\section{Classical Black Holes}

Everybody knows what the classical black hole is. In short,
black hole is a region of a space-time manifold beyond an
event horizon. In turn, an event horizon is a null surface
that separates the region from which null geodesics can
escape to infinity and that one from which they cannot.
The most general black hole is characterized by only few parameters,
its mass, angular momentum, and some conserved charge, similar to the electric
one. In general the space-time outside a black hole is stationary and axially
symmetric, but for a non-rotating black hole it is spherically symmetric and 
static. In what follows we are dealing only with spherically symmetric
black holes. 
It is important to stress that the notion of the of
the event horizon is global, it requires knowledge of both
past and future histories. To understand this better, let us consider the general
structure of spherically symmetric space-time manifolds. The metric of such 
space-time can always (at least locally) be written in the form
\begin{equation}
\label{gds}
ds^2 =  g_{00}dt^2 + 2g_{01}dtdq + g_{11}dq^2 - R^2 d\Omega^2 
\end{equation}
Four metric coefficient are functions of some time coordinate t and some radial
coordinate q, $ d\Omega =\theta^2 + \sin \theta^2 d\phi^2$ and R(t,q) is the 
radius in that sense that the sphere area is $4\pi R^2$. Making use of two
allowed coordinate (gauge) transformations it is always possible to transform
a two-dimensional (t,q) - part of the metric to the conformally flat metric.
Hence any spherically symmetric space-time is (locally) described, actually,
by only two invariant functions. One of them is, evidently, the radius $R(t,q)$.
For the other one it is convenient to choose the squared normal vector to the 
surfaces of constant radius \cite{7,8}
\begin{equation}
\label{delta}
\Delta = g^{i,j} R_{,i} R_{,j}
\end{equation}
The function $\Delta (t,q)$ brings a nontrivial information about a 
space-time structure. Indeed, in the flat Mimkowskian space-time 
$\Delta \equiv 1$, all the surfaces $R = const.$ are time-like and therefore, 
$R$ can be chosen as spatial coordinate $q = R$ on the whole manifold. 
But in the curved space-time $\Delta$ is no more constant and can in 
general be both positive and negative. The region with $\Delta > 0$ is called 
the $R$-region, and the radius can be chosen as a radial coordinate $q$. 
In the region with $\Delta < 0$ the surfaces $R = const.$ are space-like 
(the normal vector is time-like), and the radius $R$ can be chosen as a time 
coordinate $t$. Such regions are called the $T$-regions. The $R$- and 
$T$-regions were introduced by Igor Novikov.
But this is not the whole story. It is easy to show that we can not get 
$\dot R = 0$ (``dot'' means a time derivative) in a $T$-region. Hence it 
must be either $\dot R > 0$ (such a region of inevitable expansion is called 
$T_+$-region) or $\dot R < 0$ (inevitable construction, a $T_-$-region. 
The same holds for $R$-regions. They are divided in two classes, those with 
$R' > 0$ (``prime'' stands for a spatial derivative) which are called 
$R_+$-regions, and $R_-$-regions with $R' < 0$. These, $R$- and 
$T$-regions are separated by the surfaces $\Delta = 0$ which are called the 
apparent horizons. The apparent horizons can be null, time-like or space-like. 
If surface $R=0$ lies inside a $T_{-}$-region, then boundary of the latter 
forms the so-called trapped surface from which any light can escape to
the infinity. This part of the apparent horizon lies either inside a black 
hole or coincides partly with an event horizon. Clearly, the notion of the 
apparent horizon is local.

After all these preparation we consider the following example. Let us 
imagine that we have a trapped surface and,hence a black hole , and 
spherically symmetric matter layer outside it. let such a layer is held at some
fixed value of radius with the help,say,of rocket engine. If the engine is 
abruptly switched off, the layer will start to collapse and, eventually,
will cross the apparent horizon. Again a black hole will be formed, but 
now it will have different parameters (higher mass), and a new event horizon
has to be defined. In this case some part of the apparent horizon (that one 
lying in the past light cone of the crossing point ) will appear to be inside 
the new black hole, while the other part (the new one lying in the future of
the crossing point) will coincide with the new event horizon. In principle,
we may have many different matter layers, and nobody can foresee, when the 
astronauts will decide to switch off engines. Our simple gedanken experiment
demonstrates clearly the global nature of the event horizon.

All the notions introduced above are extremely important in studying
black holes formed during gravitational collapse. But there exist a rather 
simple and, nevertheless, very important model, called the eternal black
hole which we study in the next Subsection.

\subsection{Schwarzschild black hole}

All the Schwarzschild metrics (\ref{schw}) forms the one-parameter family
labeled by the value of total mass $m$, and there are no other spherically 
symmetric solutions to the vacuum Einstein equation which asymptotically flat at
infinity (this statement is known as Birkhoff's theorem). It is clear at once that 
the metric (\ref{schw}) has a singularity at the gravitational radius $r_g=2Gm$.
The nature of this singularity was not understood for a long time. And only after
works by M.Kruskal \cite{10} and I.D.Novikov \cite{9} it become clear that it is 
actually a coordinate singularity, appeared due to the static form of the line 
element (\ref{schw}) and impossibility to synchronize the clocks of static 
observers at spatial infinity ($r=\mbox{const} \to \infty$) and of the observers
(which are inevitable non-static) in the region $r<2Gm$.

The manifold, described by the line element (\ref{schw}), is not geodesically
complete. By definition, in geodesically complete space-time
(which also called a maximally analytically extended manifold) any time-like
or null geodesics should start and end either at the infinity or the singularity.
The structure of maximally extended spherically symmetric manifold very easy
to see and investigate using the so-called Carter - Penrose conformal diagrams.
On these diagrams all the infinities lie at the finite distance from the center,
and every point represents a sphere.

The conformal diagram of geodesically complete Schwarzschild space-time is shown
in Fig.1.
\begin{figure}[htbp] 
\vspace*{13pt}
\centerline{\includegraphics{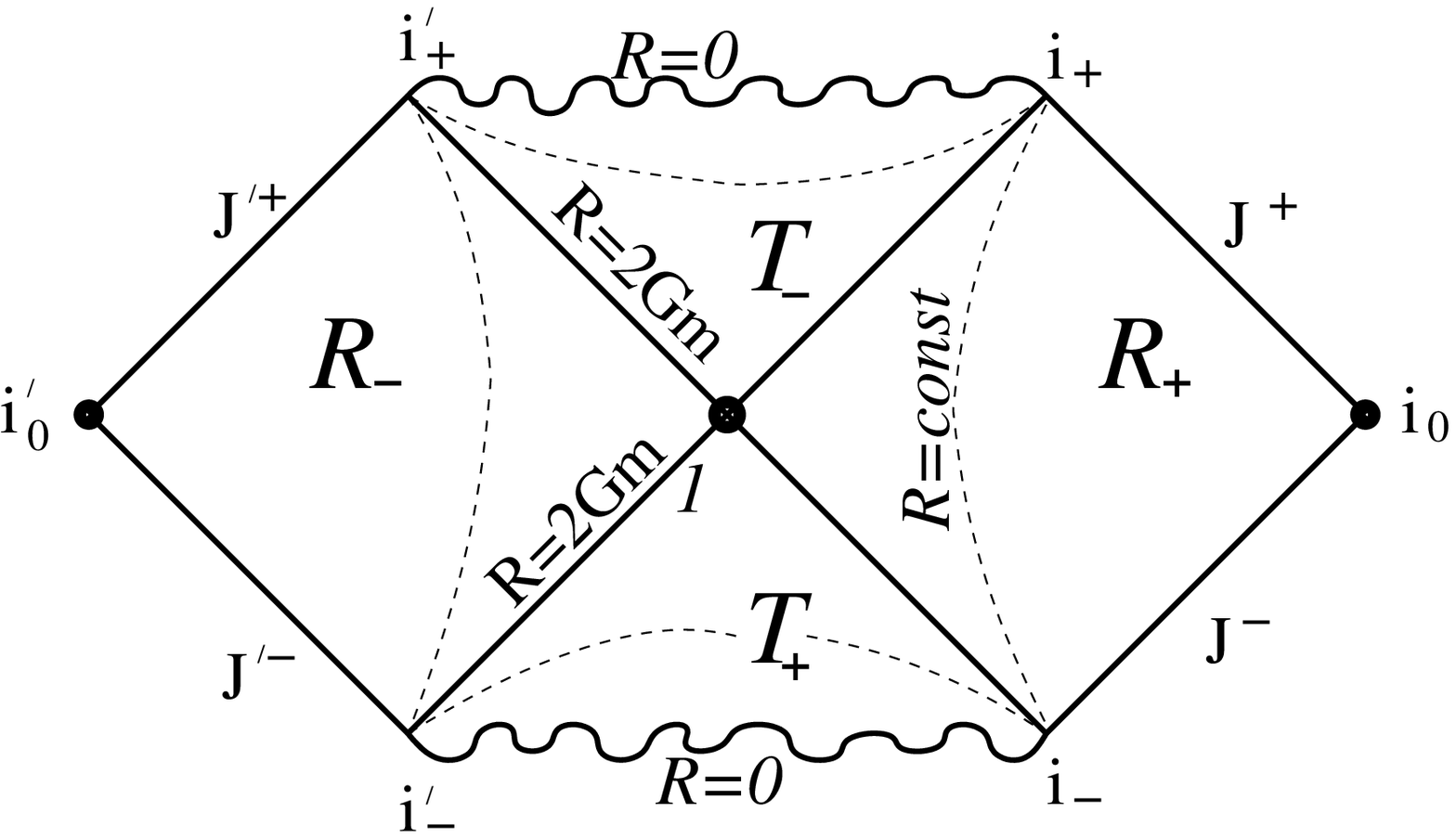}} 
\vspace*{13pt}
\caption{}
\end{figure}

We see that this manifold has a geometry of the non-traversable
wormhole. There are two asymptotically flat $R_{\pm}$-regions connected by the
Einstein-Rosen bridge. If we start from the left spatial infinity at $i^{/}_0$
and go in the direction to the right infinity at $i_0$, the radii of spheres
first decrease in the $R_{-}$-region, reach minimal value on the Einstein-Rosen
bridge and then increase in the $R_{+}$-region. The narrowest part of the bridge
called ``throat'', it lies at the intersection of two null surfaces $r=2Gm$,
which are the horizon. For the sake of simplicity we will often refer to the 
$R_{+}$-region as the 'our side', and to the $R_{-}$-region as the 'other side'
of the Einstein-Rosen bridge. The gravitating sources in the eternal Schwarzschild
black hole are assumed to be concentrated in the future and past $T$-regions
at the singular space-like surfaces of zero radius, $R=0$. These singularities are 
the real ones, because (as can be easily shown) the Riemann curvature tensor
and, hence the tidal forces, become infinite here. The Schwarzschild geometry
in the equatorial plane at the fixed moment of time can also be visualized
a geometry of some surface of rotation embedded into three-dimensional flat space.
(see Fig.2).
\begin{figure}[htbp] 
\vspace*{13pt}
\centerline{\includegraphics{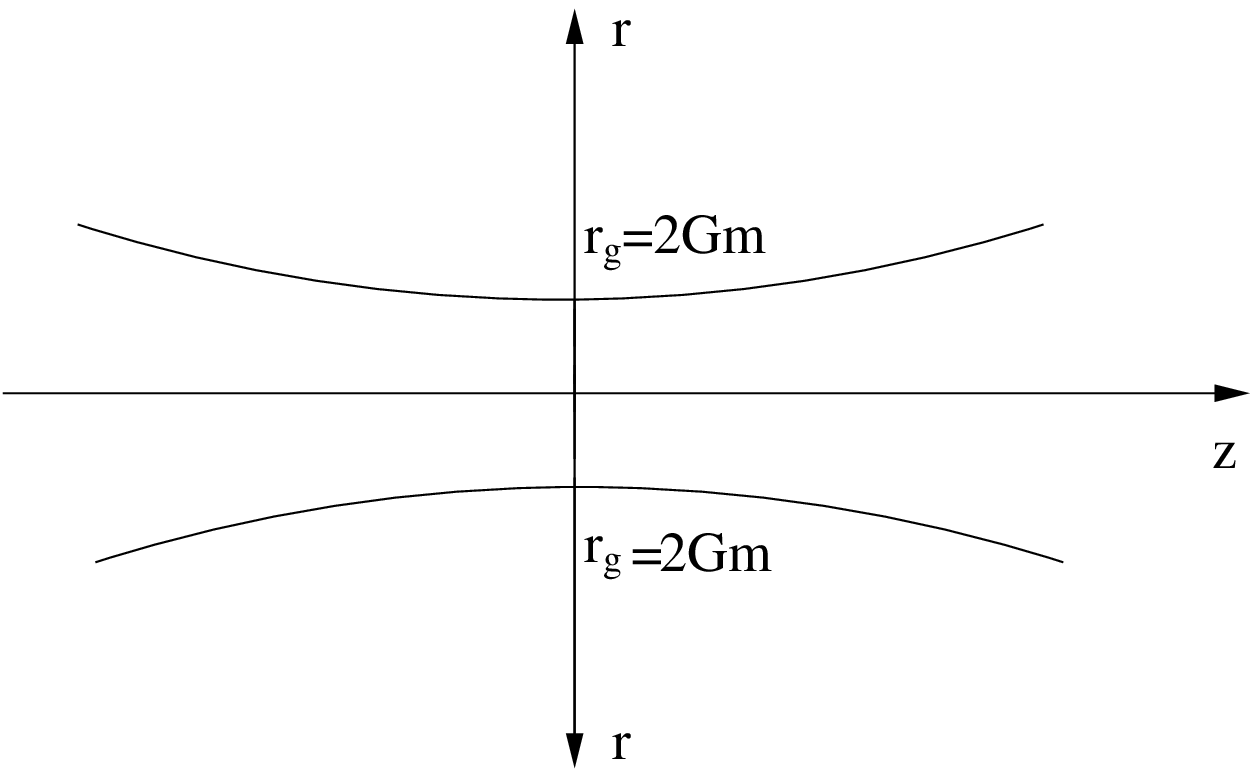}} 
\vspace*{13pt}
\caption{}
\end{figure}

In what follows we will deal sometimes with the electrically charged sources.
The space-time outside such sources is described by the Reissner-Nordstrom
line element which differs from the metric (\ref{schw}) by different function
$F$, now it becomes $F(r) = 1 - \frac{2Gm}{r} + \frac{Ge^2}{r^2}$, where $e$
is the electric charge.

\subsection{Sources}

Eternal black hole is the nice and transparent model. It appeared very useful
in investigating of classical black holes. But it really physical because
of the absence of any dynamical degrees of freedom, namely, there's no dynamics
of gravitating matter, neither real mass-shell spherically symmetric gravitons.
Therefore, having in mind the construction of quantum black hole models, we
have to include more realistic gravitating source into consideration. Because
of the nonlinearity of the Einstein equations the inclusion of source make the
quantization problem very complicated and, in general, unsolvable.
Therefore, in order to get some definite results we have to choose the most
simple types of sources. We will consider the spherically symmetric  thin
dust shell which is a direct generalization of a point gravitating mass.

The theory of thin shells in General Relativity was developed by W.Israel
\cite{11} and applied then by many authors studying problems of gravitational
collapse and cosmology \cite{12,13}. We don't need yo consider this theory
in details. In our case, both outside and inside the shell we have either
vacuum (Schwarzschild) or electro-vacuum (Reissner-Nordstrom) metric, and 
and complete information about the shell dynamics is contained in the single
equation, namely, in the $ \left( _0^0\right)$-Einstein equation for the shell
 (this is the so called energy constraint). It has the form 
\begin{equation}
\label{szero}
\sigma_{in} \sqrt{\dot\rho^2 + \Delta_{in}} - 
\sigma_{out} \sqrt{\dot\rho^2 + \Delta_{out}} = 4 \pi G \rho S_0^0 
\end{equation}
Here $\rho$ is the shell's radius as a function of the proper time 
of the observer sitting on the shell, the dot means the proper time 
derivative, $S_0^0$ is the surface energy density concentrated on the shell,
and $\Delta_{in,out}$ is the value of the (introduced above) invariant
function just inside and outside the shell. In our case it equals
$\Delta=-F$. The step function $\Delta$ take two values $\pm 1$ depending
om whether radii increase  in the direction of outward shell's normal
vector ($\sigma = +1$) or they decrease ($\sigma = -1$). It is clear
that the sign of $\sigma$ coincides with the sign of the $R$-region where
the shell moves, and its sign can be changed only in the $T$-regions.
For the sake of simplicity we consider here in details only the case when
the shell is not electrically charged and there is no other gravitating
sources inside it. Then 
\begin{equation}
\label{set}
\begin{array}{l}
\Delta_{in} = 1 ,\ \ \ \ \sigma_{in} \equiv 1\\
\\
\Delta_{out} = 1 - \frac{2Gm}{\rho} + \frac{Ge^2}{\rho^2} ,\ \ \ \ m > 0\\
\end{array}
\end{equation}
Our shell is made of dust, that is, of noninteracting (except gravitationally) 
particles. This means that th surface tension is zero, $S_2^2 = S_3^3 = 0$. 
Then, from the continuity condition, 
\begin{equation}
\label{ssdot}
\frac{dS_0^0}{d\tau} + \frac{2 \dot\rho}{\rho} (S_0^0 - S_2^2) + [T_0^n] = 0 
\end{equation} 
it follows that 
\begin{equation}
\label{M}
S_0^0 = \frac{M}{4 \pi \rho^2} 
\end{equation}
where $M$ is readily identified as a bare mass of the shell (the sum of the 
rest masses of particles without the gravitational mass defect).

Finally, we arrive at only one equation which brings all the information 
about the shell's motion, namely 
\begin{equation}
\label{emotion}
\sqrt{\dot\rho^2 + 1} - \sigma_{out} \sqrt{\dot\rho^2 + 1 - \frac{2Gm}{\rho} }
 = \frac{GM}{\rho}
\end{equation}
We can solve this equation for the total mass $m$ to get 
\begin{equation}
\label{square}
m = M \sqrt{\dot\rho^2 + 1} - \frac{GM^2}{2 \rho} 
\end{equation}
and now it is seen that it is nothing more but the energy conservation 
equation, the square root being the famous Lorentz factor written in terms 
of proper time derivatives.

We shall be interested here in bound motion only. Let us denote by $\rho_0$ 
the radius of the shell at the moment of rest, and let $\rho_0 > R_+$ 
(that is, we are outside the event horizon). Then it follows from 
Eqn.(\ref{emotion}) and Eqn.(\ref{square}) that $m < M$ and 
\begin{equation}
\label{mM}
m = M - \frac{2GM^2}{2 \rho_0}
\end{equation}

\begin{equation}
\label{sigmaout}
\sigma_{out} = Sign\left( 1 - \frac{GM}{\rho_0}\right)
\end{equation}
Differentiating Eqn.(\ref{mM}) with respect to the bare mass $M$ we obtain 
\begin{equation}
\label{dd}
\frac{\partial m}{\partial M} = 1 - \frac{GM}{\rho_0} 
\end{equation}

Let us consider the dependence of the total mass $m$ on the bare mass $M$ for 
some fixed value of the turning radius $\rho_0$. As seen form Fig.3, the 
curve has two branches, the increasing and the decreasing ones. 

\begin{figure}[htbp]
\vspace*{13pt}
\centerline{\includegraphics{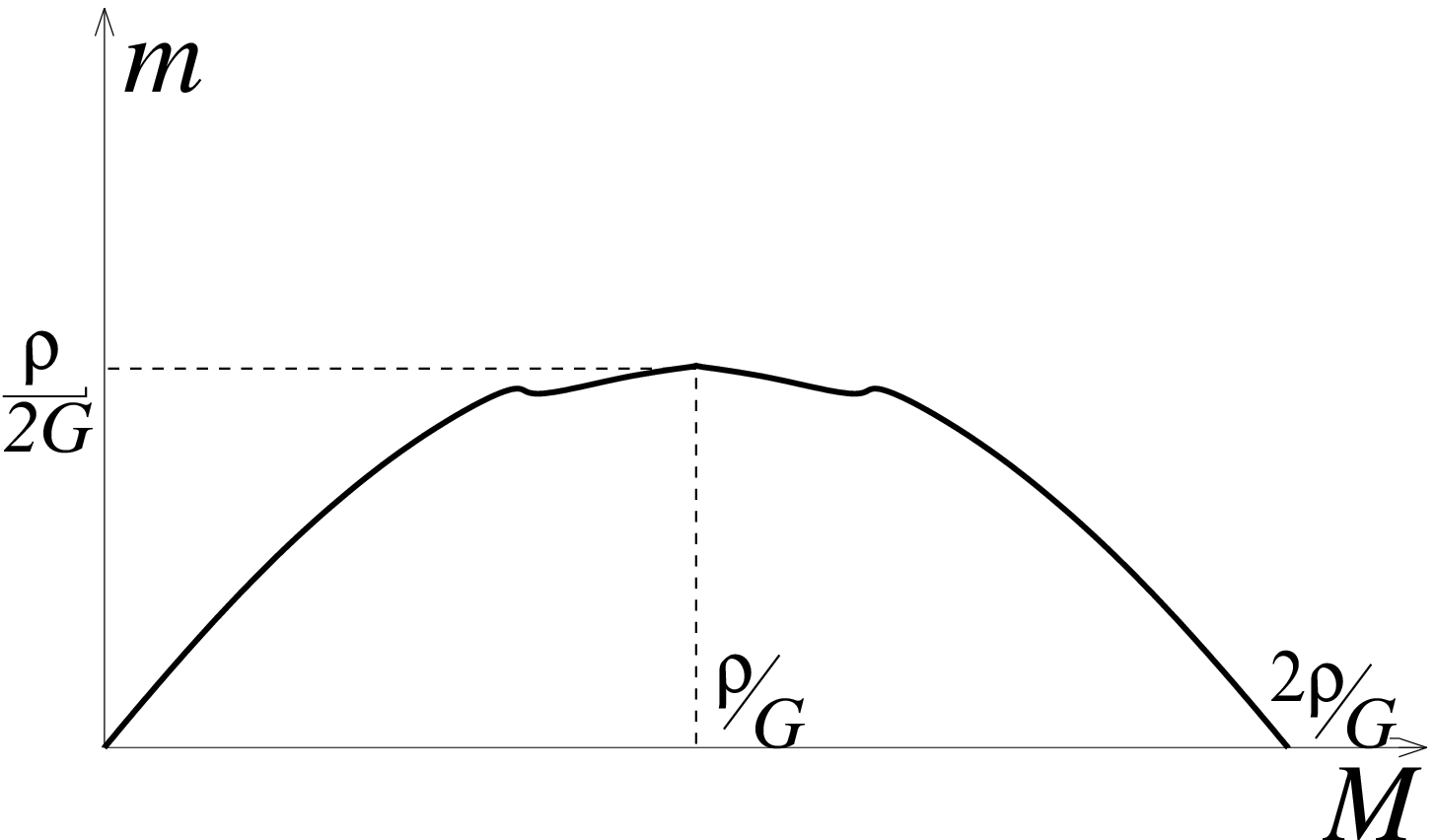}}
\vspace*{13pt}
\caption{}
\end{figure}
The total mass 
$m$ first increase with $M$, and the parameters of the shell in this case 
are such that 
\begin{equation}
\label{incr}
\begin{array}{c}
\frac{\partial m}{\partial M} > 0\\
\\
\sigma_{out} = + 1\\
\\
\frac{1}{2} < \frac{m}{M} < 1
\end{array}
\end{equation}
The world lines of these shells start from the past singularity at $R = 0$, 
pass the $R_+$-region and end at the future singularity at $R = 0$. We shall 
call such a situation the black hole case (the shell forms the black hole). 
The complete Carter-Penrose diagram is shown in Figs.4,5.
\begin{figure}[htbp] 
\vspace*{13pt}
\centerline{\includegraphics{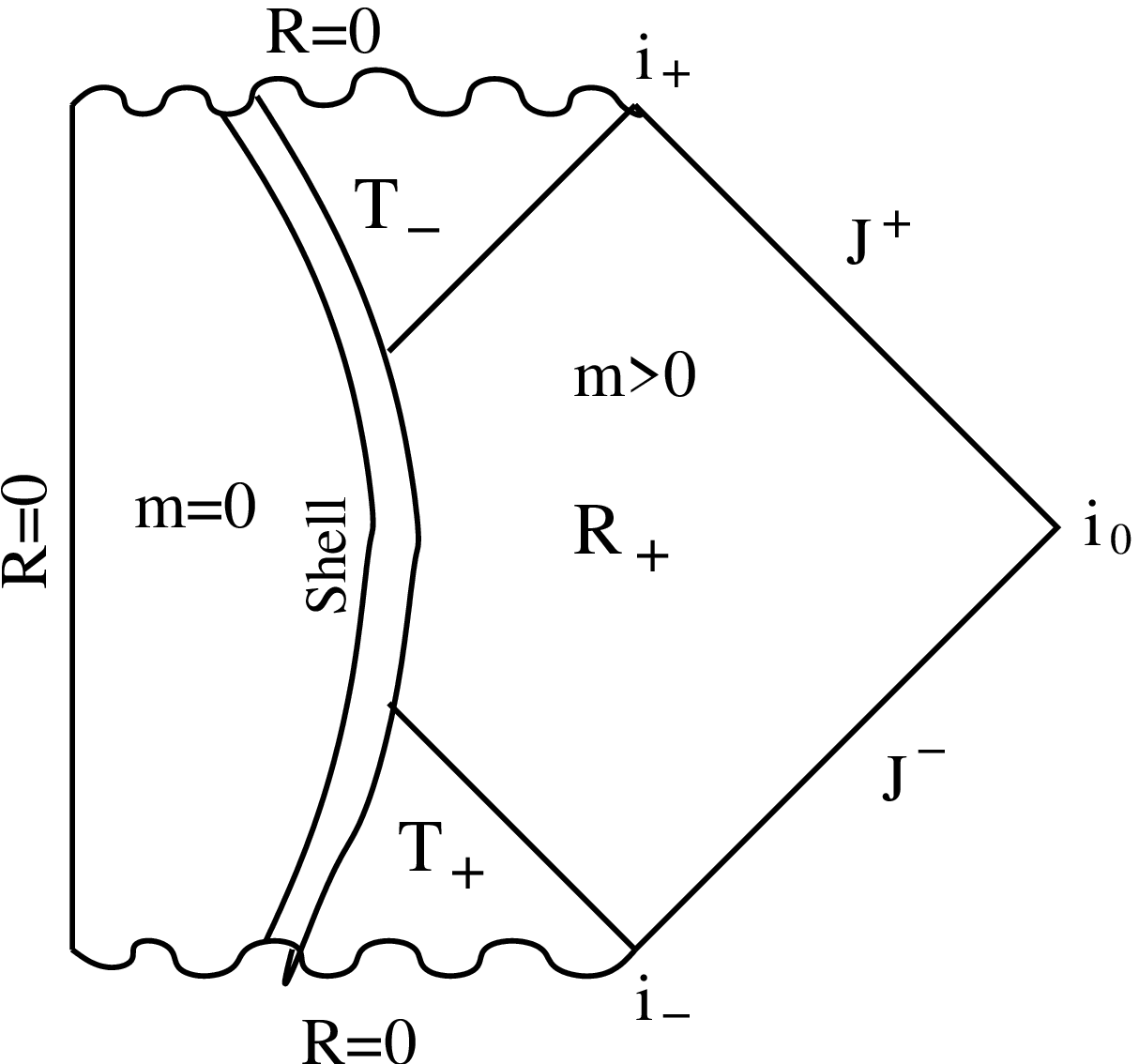}}
\vspace*{13pt}
\caption{}
\end{figure}

\begin{figure}[htbp] 
\vspace*{13pt}
\centerline{\includegraphics{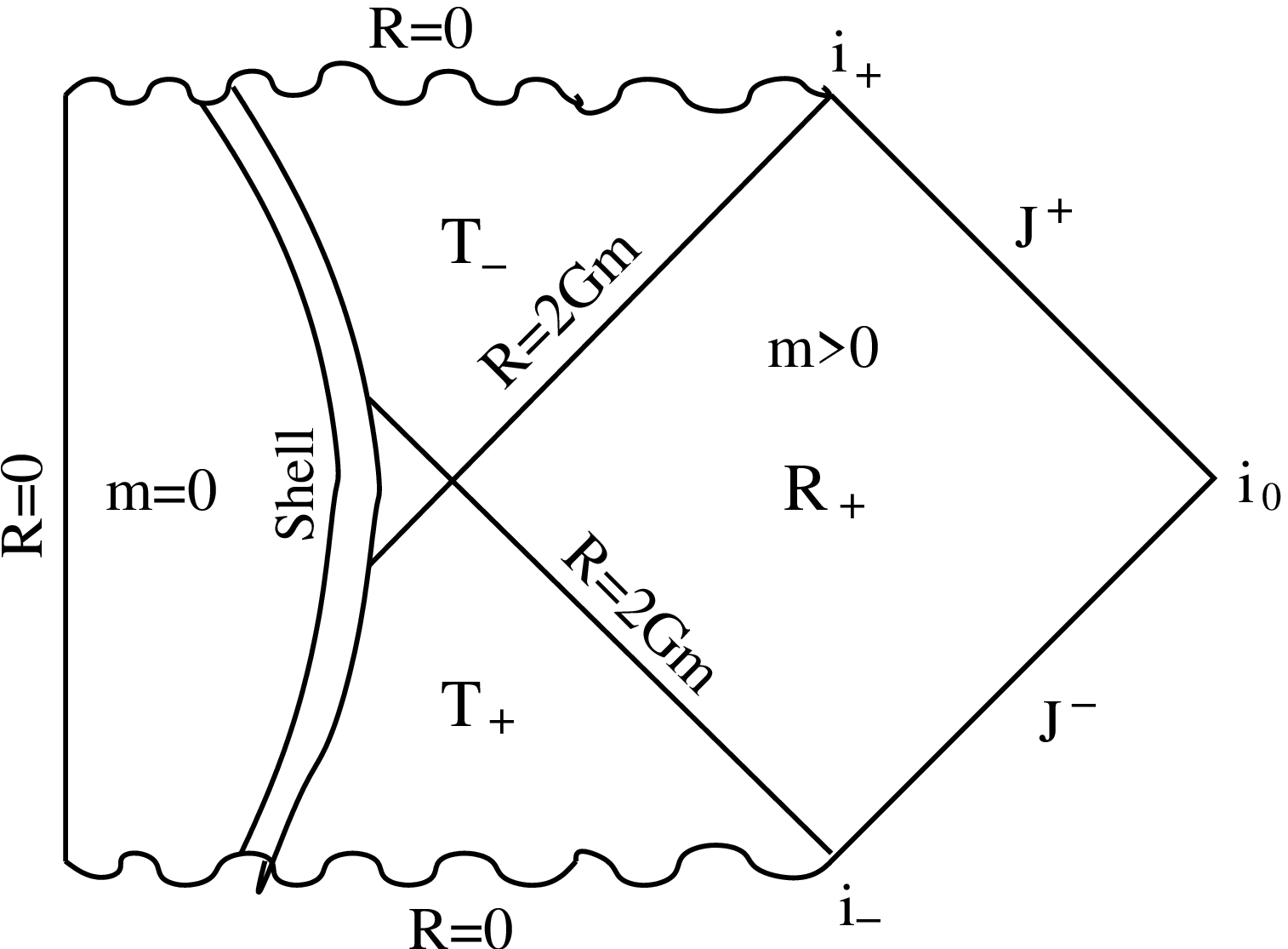}}
\vspace*{13pt}
\caption{}
\end{figure} 
The spatial geometry at the moment of time symmetry is shown in Fig.6. 
\begin{figure}[htbp]
\vspace*{13pt}
\centerline{\includegraphics{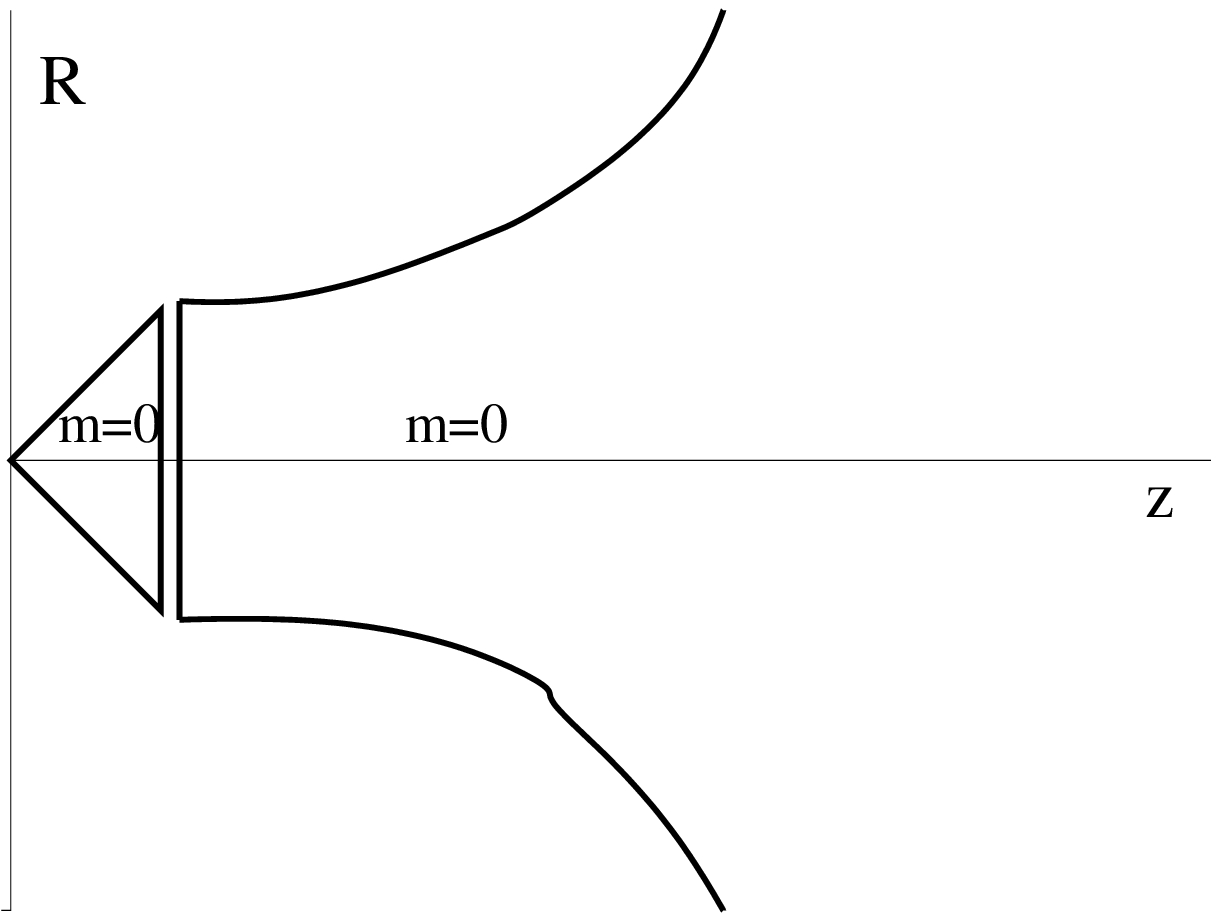}} 
\vspace*{13pt}
\caption{}
\end{figure}

\begin{figure}[htbp] 
\vspace*{13pt}
\centerline{\includegraphics{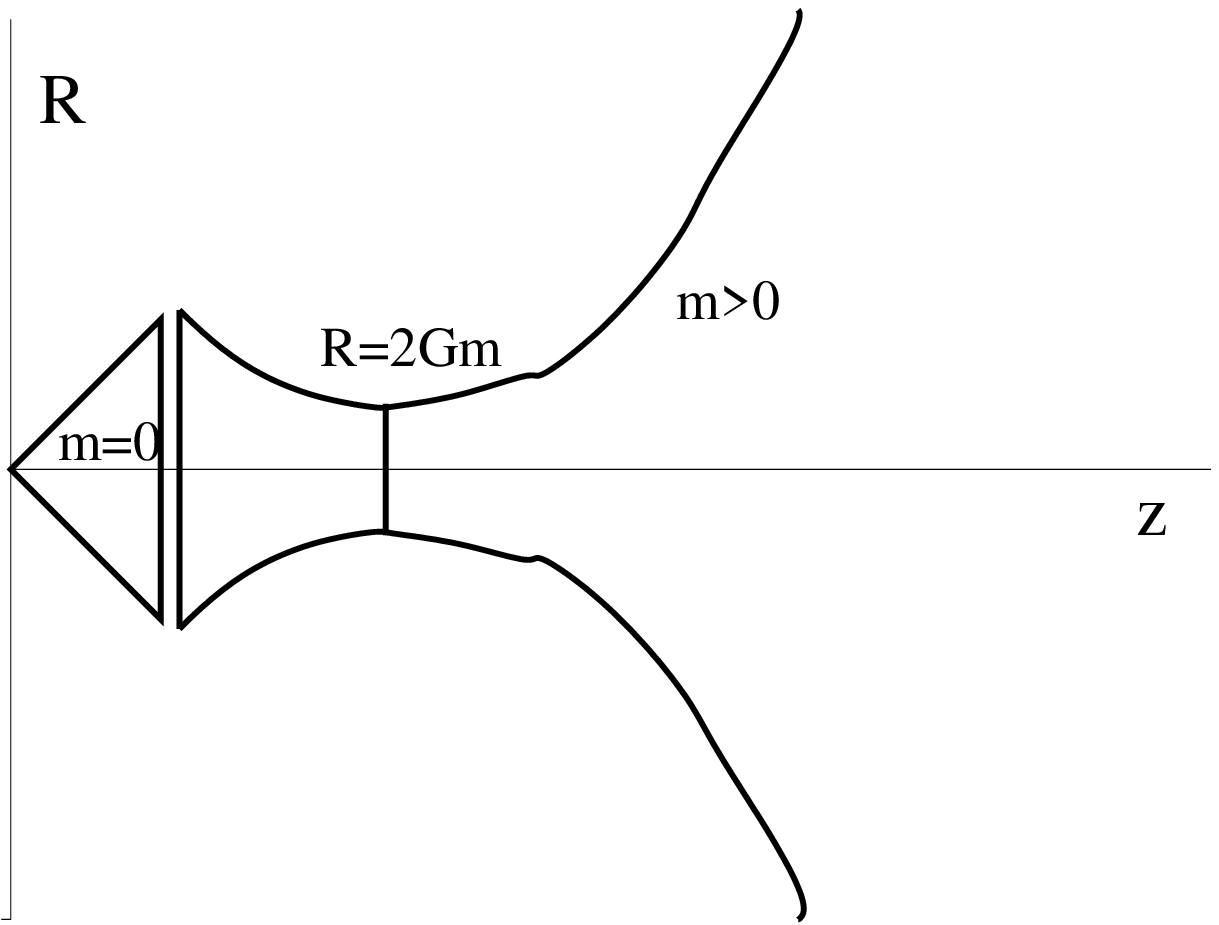}}
\vspace*{13pt}
\caption{}
\end{figure} 
The curve $m(M)$ reaches then 
the maximum and starts to decrease. For the decreasing branch we have
\begin{equation}
\label{decr}
\begin{array}{c}
\frac{\partial m}{\partial M} < 0\\
\\
\sigma_{out} = - 1\\
\\
\frac{m}{M} < \frac{1}{2}
\end{array}
\end{equation}
Now the shells pass the $R_-$-region, and we shall call this the wormhole case 
(the shell forms the wormhole on the other side of the Einstein-Rosen bridge). 
The corresponding Carter-Penrose diagrams is shown in Fig.8,9. The spatial 
geometry is shown in Fig.7. It is explicitly seen in the figure that the 
radii decrease outside the shell, reach the minimal value $2Gm$ at the 
bifurcation point and then they start to increase. The wormhole region can 
not be reached or seen from the $R_+$-region. In the marginal case $m/M = 1/2$ 
and the turning point lies exactly at the bifurcation point $R = 2Gm$. 
\begin{figure}[htbp] 
\vspace*{13pt}
\centerline{\includegraphics{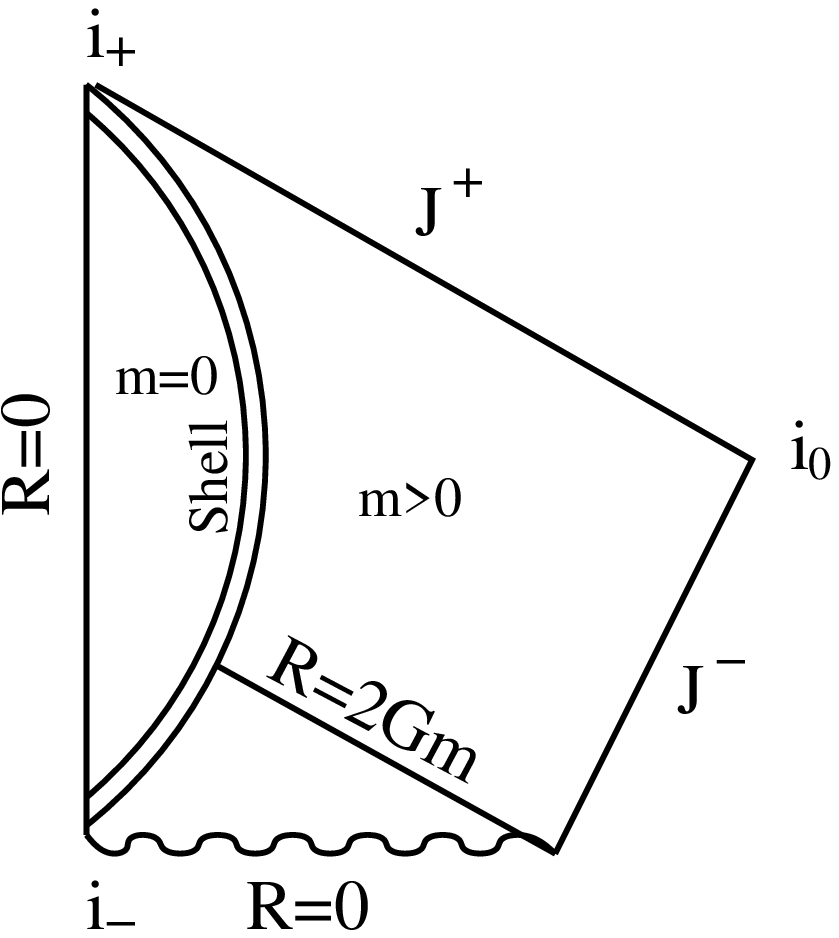}} 
\vspace*{13pt}
\caption{}
\end{figure}

\begin{figure}[htbp]
\vspace*{13pt}
\centerline{\includegraphics{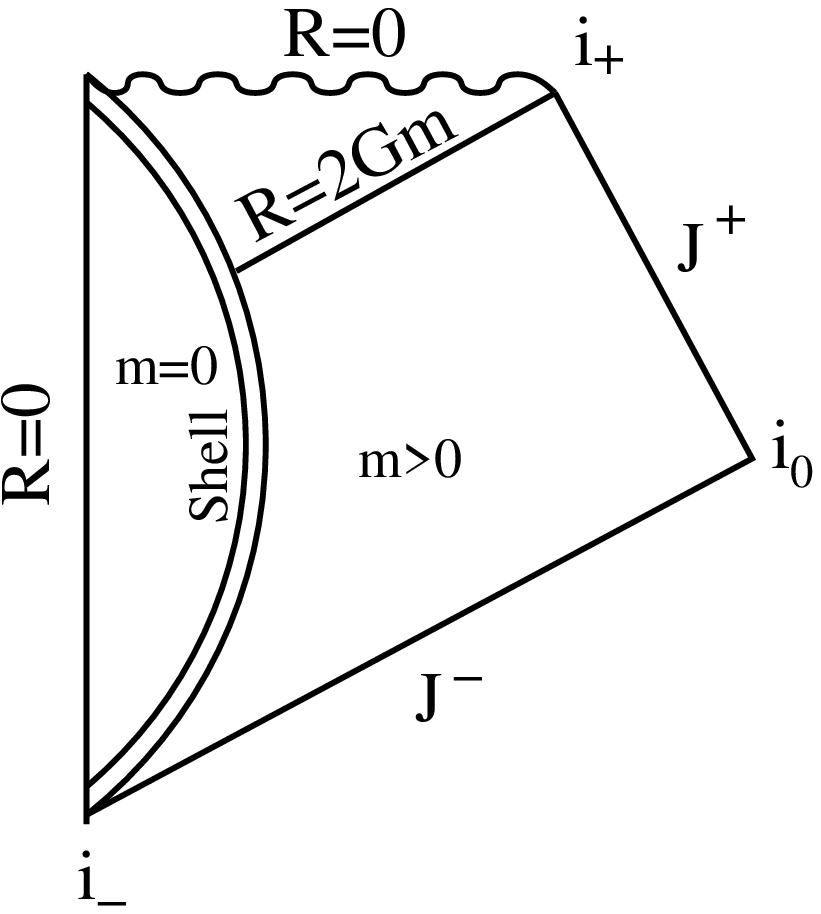}}
\vspace*{13pt}
\caption{}
\end{figure} 

It is instructive to consider some limiting cases. The first one is the 
limit $M \to \infty$. When the bare mass is increasing and 
$\frac{m}{M}<\frac{1}{2}$, the turning point is further and further from
the horizon, moving toward the ``left'' infinity (on the other side 
Einstein-Rosen bridge). And as a limit we get geometry of the eternal black
hole everywhere except at the left infinity which is now singular.

Another limiting case is a completely closed world with zero total mass 
$m = 0$. We have for such a situation
\begin{equation}
\label{mzero}
\begin{array}{c}
m = 0\\
\\
M = 2 \rho_0/G
\end{array}
\end{equation}

For completeness let us consider the thin shell with negative total mass 
$m < 0$ (but with positive bare mass $M$). From Eqn.(\ref{emotion}) it 
follows immediately that $\sigma_{out} = - 1$ everywhere, the radii thus 
decrease outside the shell up to the singularity at $R = 0$. The space-time 
topology is $S^3 \times R^3$ rather than $R^3 \times R^1$ as it is for 
positive total masses $m$. So, in this case we have no spatial infinity at 
all. In what follows we will consider only positive total masses.

\subsection{Black hole universality}

It was already mentioned  that the most general black hole is the stationary
axially symmetric space-time manifold, describing by a very limited set
of parameters, namely, by its mass, angular momentum and (locally)
conserved charges. But what does happen when the initial state of the 
collapsing matter is not exactly spherically symmetric (here we consider
the case of zero angular momentum)? The huge work of many researchers
has led to the following result (the detailed review of this topic can be
found in the book \cite{14}). In the process of black hole formation
everything that can be radiated is radiating away.
It concerns all the multipoles (including the  monopole moment) of scalar
fields, the higher multipoles of electromagnetic, nonabelian and 
gravitational fields. In some theories the scalar monopole can not be
radiated away, but only in those cases when it is rigidly linked to the 
monopole moment (Coulomb part) of the corresponding nonabelian field.
This feature of process gravitational collapse was first understood by
J.A.Wheeler and formulated as the (now famous) statement that
``The black hole has no hair''. Thus, the process of black hole formation
is accompanied, in general, by the loss of information about the initial
state and the preceding history of collapsing matter. 

From the very definition of the black hole it is clear that nothing can go
off the event horizon. But this does not mean automatically that the 
black hole mass cannot be  decreased. The matter is that in case of 
nonzero angular momentum there exist around the event horizon a special
region, called ergosphere, inside which there can be states with negative
energy (from the point of view of a distant observer). Then, the following 
process can take place. The particles that fall into the ergosphere, may
decay into pairs, and one of the component whose energy is negative, may
fall into the black hole, while another one would go out of the ergosphere
(but not out of the black hole!) and escape to the infinity, thus 
decreasing the black hole mass. In the literature it is known as the 
``Penrose process''. If instead of particles, we are dealing with the fields,
then the analogous process also will take place, it is called 
``superradiance''. Both Penrose process and superradiance result in
decreasing the black hole angular momentum. And the real universality of
black holes is that fact that it is impossible by any classical process
to decrease the area of the black ole boundary, i.e. for the event horizon
area $A$ we have always
\begin{equation}
\label{A}
d A \ge 0 
\end{equation}
\newpage

\section{Quantum Era of Black Hole Physics }

\subsection{Black hole thermodynamics}

In this section (and only in this) subsection we will use some formulas for 
a general axisymmetric black hole with total mass $m$, electric charge $Q$
and angular momentum $J$. For such a black hole 
Being the final state of the gravitational collapse, black holes have very 
interesting properties.It appeared that black holes, during their formation, 
emit everything that can be emitted. Thus, the black hole resembles the body 
in thermodynamical 
equilibrium. In 1972 Jacob Bekenstein suggested that this is not merely a 
coincidence, and the black hole really brings some amount of entropy and has 
some definite temperature. This point of view was supported by the following 
inequality already known at the time, namely, that the area of the black hole 
horizon can not decrease.This resembles the second law of thermodynamics. 
J.Bekenstein suggested that 
the black hole entropy is proportional to the black hole area $A$ (more 
general, the entropy may be equal to some increasing function of the area).  

In this section (and only in this) subsection we will use some formulas for 
a general axisymmetric black hole with total mass $m$, electric charge $Q$
and angular momentum $J$. For such a black hole
\begin{eqnarray}
\label{ang}
A = 4 \pi \left(R_+^2 + \frac{J^2}{m^2} \right), \nonumber \\
\\
R_+ = G\left( m+ \sqrt{m^2 - \frac{Q^2}{G} - \frac{J^2}{m^2G}}\right)
\end{eqnarray}
 Moreover, it appeared that the first law of thermodynamics is also valid 
and can be expressed by the following mass formula.
\begin{equation}
\label{mass}
\delta m = \frac{\kappa}{8 \pi} \delta A + \Omega \delta J + \Phi \delta Q
\end{equation}
Here $\delta m$ is the mass difference between two stationary black holes 
with slightly different areas $\delta A$, angular momenta $\delta J$ and 
electric charges $\delta Q$, while $\Omega$ is the black hole's angular 
velocity and $\Phi$ is the value of the Coulomb potential at its surface. 
The prefactor $\kappa$ in front of $\delta A$ is the so called surface 
gravity which equals to 
\begin{equation}
\label{kappa}
\kappa = \frac{4 \pi}{A} \sqrt{M^2 - Q^2/G - J^2/M^2 G}
\end{equation}
Following J.Bekenstein, the surface gravity should be proportional to the 
black hole temperature $\Theta$, $\kappa=\alpha \Theta$.

Let us consider a Schwarzschild black hole as a thermodynamical system and 
put it in a thermal bath at some fixed temperature. Then, if the temperature 
of the black hole is the same as that of the thermal bath, the amounts of 
the absorbing and emitting radiation are equal, the black hole and the bath 
are in thermal equilibrium. But this equilibrium is unstable. The reason for 
this is the negative heat capacity of the Schwarzschild black hole. Indeed, 
from the definition of the heat capacity we have
\begin{equation}
\label{heat}
C = \Theta \frac{\partial S}{\partial \Theta} = - 8 \pi G m^2 < 0
\end{equation}
Let us suppose that the black hole is initially in thermal equilibrium with 
the environment. Then, if by some fluctuation the mass of the black hole will 
become a little bit larger than its equilibrium value the temperature will 
become lower than tat of the thermal bath causing the absorption of 
radiation, further increase of the mass and decrease of the temperature, 
thus resulting in the unbound growth of the black hole size and mass. 
If the initial fluctuation results in the small decrease of the mass and, 
consequently, in the increase of the black hole temperature, the black hole 
starts to emit radiation more than to absorb it, thus further decreasing the 
mass up to the very end. In this case the black hole disappears completely. 

In the case of the Reissner-Nordstrom black hole the situation is different. 
The temperature is zero for the extreme black hole $(m = |Q|/G)$, reaches its 
maximum and then decreases to zero when the mass infinitely increases. 
The heat capacity is positive near the extreme state, first grows to infinity, 
then becomes negative like in the Schwarzschild case, We see that the charged 
nearly extreme black hole is thermally stable like the ordinary bodies. 

\subsection{Hawking radiation}

The quantum era of the black hole physics began in 1974 with Hawking's 
discovery of the black hole evaporation. Stephen Hawking considered quantum 
vacuum fluctuations of matter fields on the Schwarzschild black hole 
background and showed that the very existence of the event horizon leads to the 
radiation flow from the black hole (violating the classical theorem 
$dA \ge 0$). The surprising fact is that this radiation is nothing more but 
the blackbody radiation with the temperature 
\begin{equation}
\label{temp}
\Theta = \frac{\kappa}{2 \pi}
\end{equation}
Thus, the black hole entropy is exactly one fourth of the dimensionless 
horizon area
\begin{equation}
\label{entropy}
S_{BH} = \frac{1}{4} \frac{A}{l_{pl}^2}
\end{equation}
which is in full agreement with the Beckenstein's conjecture (we use the 
units in which the Planckian constant $\hbar$, the light velocity $c$ and the 
Boltzmann constant $k$ are put equal to one; Planckian length equal to 
$l_{pl} = \sqrt{\hbar G/c^3} = G^{1/2}$, and Planckian mass is 
$m_{pl} = \sqrt{\hbar c/G} = G^{- 1/2}$). 

Remarkably enough the same value 
of the black hole temperature can be obtained in a quite different way.
By making a Wick rotation of the $M_2$-part of the Schwarzschild or 
Reissner-Nordstrom line element written in the curvature coordinates we 
obtain a two-dimensional Euclidean half-plane 
($-\infty < t_E < \infty, R_+ < R < \infty$). If we place the origin at 
$R = R_+$, assume periodicity of the Euclidean time $t_E$ and identify 
points over the period, we then get a two-dimensional surface with, in 
general, a conical singularity at the origin. This singularity can be removed 
by a suitable choice of the time period. And the inverse of such a period
is equal to the temperature in finite temperature quantum field
theories and in our case it can be called a topological temperature.
Practically, it can be done as follows. The line element of the two-dimensional 
Euclidean surface is 
\begin{eqnarray}
\label{euc}
dl^2 = F d\tau^2 + \frac{1}{F} dr^2 , \nonumber \\
\\ 
F = 1 - \frac{2 G m}{r} + \frac{G e^2}{r^2} = 
\frac{1}{r^2} (r - r_+)(r - r_-) , \nonumber \\
\\
r_{\pm} = G \left (m \pm \sqrt{m^2 - \frac{e^2}{G}} \right ) .
\end{eqnarray} 
The Euclidean time $\tau$ is assumed to be a cyclic coordinate. In the vicinity of the 
horizon $F = (r - r_+)(r - r_{-})/r_+^2$ . 
Introducing a new coordinate $\rho = 2 r_+ \sqrt{r - r_+}/\sqrt{r_+ - r_-}$, 
we get
\begin{equation} 
dl^2 = \frac{(r_{+} - r_{-})^2}{4 r_+^4} \rho^2 d\tau^2 + d\rho^2 = 
d\rho^2 + \rho^2 d\varphi^2 .
\end{equation}
And this is nothing more but the metric of a locally flat two-dimensional surface, 
written in the cylindrical coordinates. The point $\rho = 0$ is, in general, 
a conical singularity. To remove it, we have to require the period of the 
azimuthal angle $\varphi$ to be $2\pi$. Thus, 
\begin{equation}
\label{prn}
T = \frac{4 \pi r_+^2}{r_{+} - r_{-}} .
\end{equation}
Its inverse, 
\begin{equation}
\label{trn}
\Theta = \frac{2 G \sqrt{m^2 - \frac{e^2}{G}}}{A}
\end{equation}
equals exactly the temperature of the Reissner-Nordstrom black hole. 

All this means that quantum matter fields in the black hole background are described, 
in fact, by the finite temperature quantum field theory. In this sense, the appearance 
of the black hole temperature looks quite naturally. 

The violating of the classical law $d A \ge 0$ does not mean that the radiation 
can be emitted off the black hole, because the black hole itself is still considered 
as the classical object. The calculations of the stress-energy tensor of the 
vacuum fluctuations show that their energy density is negative. Therefore, the negative 
energy flow falls into black hole, thus diminishing its mass, while the, equal to it, 
positive energy flow escapes to infinity. One can consider such a situation as that 
the quantum fluctuations of matter fields form something like an ergosphere in the 
vicinity of the event horizon, and inside this ergosphere the Penrose process 
(for particles) or the superradiance effect (for radiating fields) may take place. 

\subsection{Rindler's space-time.}

Except the black holes there is one more example of the space-time manifold which observers 
``see'' a black body radiation. This is the Rindler's space-time. It is locally flat, 
therefore, sufficiently simple and all the quantum field theory equations can be 
explicitly solved and investigated in details. On the other hand, this example is very 
important since there is exists a deep relation between Rindler's and black hole manifolds. 
Let us consider the two-dimensional Minkowski space-time with the metric 
\begin{equation}
\label{m}
ds^2 = dt^2 - dx^2. 
\end{equation}
The Rindler's coordinates are obtained by the following transformation  
\begin{eqnarray}
\label{cr}
t = \pm \frac{1}{a} e^{a\xi} \sinh{a\eta} \nonumber \\
\\
x = \pm \frac{1}{a} e^{a\xi} \cosh{a\eta} , 
\end{eqnarray}
The upper (lower) sign is for $x >(<) 0$. The line element now reads as follows 
\begin{equation}
\label{r}
ds^2 = e^{2a\xi} (d\eta^2 - d\xi^2)
\end{equation} 
Rindler's observer, sitting at $\xi = const.$, experience a constant proper acceleration 
$\alpha = a e^{-a\xi}$. Being locally flat, the Rindler's manifold differs, nevertheless, 
from the Minkowski manifold, it has different boundaries. The boundary of the Minkowski 
space time are trivial infinities (spatial, temporal and null). Rindler's coordinates 
cover only one half of the Minkowski  space-time, corresponding to $|x| \ge |t|$, and 
the lines $x = \pm t$ are the event horizons that serve as new boundaries (in addition 
to the infinities). Exact quantum field theory calculations show (the details can be 
found in the book \cite{14a}) that the Rindler's vacuum is, actually, with the black 
body radiation which temperature equals the so called Unruh's temperature 
$T_0 = \frac{a}{2\pi}$. Every local observer measures its own local temperature 
$T = T_0 (g_{00})^{-1/2} = T_0 e^{-a\xi} = \frac{\alpha}{2\pi}$. 

It should be stressed that the appearance of the temperature (and heat bath)
in Rindler's space-time is a global effect, depending essentially on the 
boundary condition at the event horizon. And we can do here the same trick
as we did with the black hole space-time. Namely, we can introduce the cyclic
imaginary time $\eta \to i\eta = \tau$ and spatial coordinate 
$\zeta = \frac{1}{a} e^{a\xi}$ the line Rindler's element now becomes
\begin{equation}
\label{re}
dl^2 = a^2 \zeta^2 d\tau^2 + d\zeta^2 . 
\end{equation}
The obtained two-dimensional euclidean surface has, in general, a conical 
singularity that disappeared only if the period of our imaginary time $\tau$
equals $\frac{2\pi}{a}$. The inverse to this period is exactly the Unruh's
temperature. Note,that procedure, described above, is of global character,
because the choice of the period in the point dictates actually the period
of the ``proper'' imaginary time for all Rindler's observers.

Let us consider now a static spherically symmetric space-time the line element
of which can always be written in the form
\begin{equation}
\label{st}
ds^2 = e^{\nu} dt^2 - e^{\lambda} dr^2 -r^2 d\Omega^2 , 
\end{equation}
where  $\nu$ and  $\lambda$ are function of the radius $r$. By the equivalence
principle, the static observer experiences a constant acceleration. And indeed,
it can be easily shown by direct calculation that the acceleration equals
$\alpha = \kappa e^{-\nu /2}$. where

\begin{equation}
\label{kappa}
\kappa = \frac{\nu^{\prime}}{2} e^{\frac{\nu - \lambda}{2}} 
\end{equation} 
is the so called surface gravity. A static observer, sitting at $r = r_0 \ne 0$
can forget about the angular part of metric (\ref{st}) and consider a 
2-dimensional static curved space-time. such a 2-dimensional surface can be 
easily transformed to the local Rindler-like coordinates. Indeed,
\begin{equation} 
\label{rl}
ds^2 = e^{\nu} dt^2 - e^{\lambda} dr^2 = e^{\nu} ( dt^2 
- e^{\lambda - \nu} dr^2) = e^{\nu} (d\eta^2 - d\xi^2) , 
\end{equation}
where $\eta = t$ and $d\xi = e^{\frac{\lambda - \nu}{2}}$. We can calculate
the acceleration parameter $a$ in the following way. Let 
$r = r_0 + \delta r$, then
$\delta\xi = e^{\frac{\lambda - \nu}{2}} \delta r$
The conformal factor near $r = r_0$ equals 
$e^{\nu}(r_0)(1 + \nu^{\prime} \delta r) = 
e^{2 a \xi_0} ( 1 + 2 a \delta\xi)$.
From this it follows that $e^{2 a \xi_0} = e^{\nu}(r_0)$ and
$a = \frac{\nu^{\prime}}{2} e^{\frac{\nu - \lambda}{2}}(r_0)$
which is exactly the surface gravity $\kappa$

The equivalence of the Rindler's observers to the static ones in the spherically 
symmetric space-times does not mean that the latter is heated. This
equivalence is only local. To have the global equivalence (of course, with 
different Rindler's observers for different static observers) we need the
event horizon, and that is exactly what we have in the case of spherically
symmetric black holes. But this is not the end of the story. The reduction of
the four-dimensional spherically symmetric space-time to 2-dimensional
$(t,r)$-surface can be safely done only if $r \ne 0$. The point $r = 0$ is
the coordinate singularity. Thus, the equivalence described above is valid
only if the corresponding euclidean surface does not contain such a point.
Otherwise we would have to consider spherically symmetric Rindler's manifold,
which essentially non-static.

And,finally, the last note. The temperature (if it exist) measured by a
local static observer obeys the law $\Theta \sqrt{g_{00}} = \mbox{const}$.
Therefore, the temperature $\Theta_0 = \frac{\kappa}{2 \pi}$ is measured by
observer, sitting at the point where $g_{00} = 1$. For the asymptotically
flat space-time this is, of course, the point at spatial infinity.

\subsection{Quantum mass spectrum}

Historically, the discovery of the quantum nature of radiation has led to the 
construction of the hydrogen atom model and to the explanation of its discrete 
energy spectrum. Analogously, the quantum nature of radiation suggests that 
the mass spectrum of black holes is, actually, quantized. Jacob Bekenstein 
proposed the equidistant spectrum for the black hole horizon area. His 
motivation is as follows. The entropy value is just the logarithm of the 
number of ways in which we can construct the black hole with given parameters 
and related to the amount of information hidden inside the black hole. 
The amount of information is naturally quantized. Therefore, the black hole 
area which is proportional to the entropy should have an equidistant spectrum. 
A more quantitative explanation is due to Slava Mukhanov. We generalized 
his result originally derived for the Schwarzschild black hole, to the 
Reissner-Nordstrom charged black hole. Let the black hole 
emits the quantum energy $\omega$ due to transition from the $n$-th state 
to the $(n-1)$-th state. The ``typical'' energy of this quantum minus 
the work done against the Coulomb attraction should be proportional to the 
temperature, i.e.
\begin{equation}
\omega_{n,n-1} - \Phi dQ = \pi\alpha\Theta ,
\label{omega}
\end{equation} 
where $dQ$ is the charge of the quantum in question, and $\alpha$ is a 
slowly varying function of the integer quantum number $n$. Inserting, then, 
Eqn.(\ref{omega}) into Eqn.(\ref{mass}) we have $(\omega_{n,n-1}=dm)$
\begin{equation}
dA = 4\pi G \alpha ,
\end{equation}
or
\begin{equation}
dA = 4\pi G \alpha dn ,
\label{dA}
\end{equation}
Integrating we get
\begin{equation}
A = 4\pi G \tilde\alpha n + 4\pi G C ,
\label{area}
\end{equation}
where $C$ is a constant of integration, and $\tilde\alpha$ is some another slowly varying 
function of $n$. It is natural to suppose that the value $n = 0$ 
corresponds to the minimal possible value of the horizon area, e.i., it 
should correspond to the extreme black hole. This gives us $C = Q^2$. 
It is clear that the Hawking's formulae are valid only if the back reaction 
of the radiation on the space-time metric can be neglected, e.i., for law 
(comparing to the black hole mass) temperature. Thus, we are interested in 
two different cases, in the large black hole regime and in the nearly 
extreme regime.  
In the first case $n\gg Q^2$ and we assume the following expansion
\begin{eqnarray}
\tilde\alpha = \alpha_1 + \beta_1\frac{Q^2}{n} + \gamma_1\frac{Q^4}{n^2} + 
\ldots ,\\
\nonumber\\
\frac{Q^2}{n} \ll 1 ,\nonumber
\label{ll}
\end{eqnarray}
while in the second case $n\ll Q^2$ and we have
\begin{eqnarray}
\tilde\alpha = \alpha_2 + \beta_2\frac{n}{Q^2} + \gamma_2\frac{n^2}{Q^4} + 
\ldots ,\\
\nonumber\\
\frac{Q^2}{n} \gg 1 .\nonumber
\label{gg}
\end{eqnarray}
The corresponding expansions for the black hole mass are
\begin{eqnarray}
m &=& \frac{\sqrt{n}}{2\sqrt{G}}\sqrt{\alpha_1} \left(1 + 
\frac{\beta_1+3}{2\alpha_1} \frac{Q^2}{n} \right.
\nonumber \\
&+& 
\left. \left(\frac{\gamma_1}{2\alpha_1} - 
\frac{(\beta_1+1)(\beta_1+4)}{8\alpha_1^2}\right)\frac{Q^4}{n^2} +
\ldots\right) ,\\
\nonumber\\
&&n \gg Q^2\nonumber
\label{mll}
\end{eqnarray}
and
\begin{eqnarray}
m = \frac{|Q|}{2\sqrt{G}} \left(2 + \frac{\alpha_2^2}{4} \frac{n^2}{Q^4} +
\ldots\right) ,\\
\nonumber\\
&&n \ll Q^2\nonumber
\label{mgg}
\end{eqnarray}
The values of $\alpha$'s, $\beta$'s and $\gamma$'s can and will be determined 
in the next Section for the specific model.

\section{Quantum Black Hole Models}

\subsection{Eternal black hole}

We have already described in detail the classical eternal black hole and 
mentioned that it has no dynamical degrees of freedom. In this respect it is
analogous to the Coulomb field in electrodynamics. Nevertheless it is 
instructive to quantize such a system. Of course, the full program of
quantization can not be fulfilled because quantum gravity is not renormalizable
pertubatively. But the quantization of the vacuum spherically symmetric
space-times can be done in a nonperturbative way in the 
so-called frozen formalism.
In this formalism we quantize only one, radial, degree of freedom. Though
not complete, such a procedure can give us rather reliable results, especially
in the light of the recent result \cite{16} which support the suggestion that
the four-dimensional Einstein gravity may appear nonperturbatively 
renormalisable \cite{17}. Quantization of the eternal black hole in frozen
formalism was done by H.Kastrup and T.Thiemann \cite{18} in the so-called
quantum loop gravity and by K.Kuchar \cite{19} who used the conventional
geometrodynamics. Even for such a simple space-time manifold the problem
appeared nontrivial. But the final result was the same in both approaches,
namely, the quantum functional on spherically symmetric vacuum metrics
depends only on the Schwarzschild mass. And in the absence of dynamical
degrees of freedom it is impossible to extract any information about a black
hole spectrum.
Physically such a trivial result can be understood if one compares  the eternal
black hole with its electro-dynamical analog, the hydrogen atom. Let us imagine
that we first allow the electron to ``fall'' classically on the proton and then
start quantization procedure. What we would get? Absolutely nothing. Of course,
in electrodynamics we have both positive and negative charges, and the mass
of the eternal black hole can only be positive. And, instead of ``nothing''
the curvature singularity is developing in the process of gravitational
collapse. But the qualitatively, the picture is the same. Therefore in order
to obtain physically meaningful  results we have to ``switch on'' 
some dynamical
degrees of freedom. In subsequent Subsection we consider the quantization
of spherically symmetric thin dust shells, for which we already developed
the classical theory.

\subsection{Naive quantization}

The material of this section is based on the works \cite{20,21,22}.
Now, we investigate a specific quantum black hole model, which 
classical counterpart is described in the first part of the lectures. 
It is a self-gravitating spherically symmetric thin dust shell with bare 
mass $M$, total mass $m$ and electric charge $e$. The ``correct'' quantization 
procedure (which starts from the ``first principles'') we postpone till 
the next Chapter and consider here a very simple quantum model based on 
the classical equation of motion for our thin shell. Since $m$ is a 
total mass of the system and the Eqn.(\ref{square}) represents the 
energy conservation law it is reasonable to consider it equal numerically 
to the Hamiltonian of the system in question. Thus, we have 
\begin{equation}
\label{square1}
m = M \sqrt{\dot\rho^2 + 1} - \frac{GM^2 - e^2}{2 \rho} 
\end{equation}
  
To quantize a system we should construct first the classical Hamiltonian 
and then to replace the classical Poisson brackets by the quantum 
commutation relations. Given the classical expression $E(\dot\rho, \rho)$ 
for the energy of the system as a function of its radius and velocity 
(actually, rapidity in our case), the conjugate momentum $p$, 
Lagrangian $L$ and Hamiltonian $H$ can be found from the following relations
\begin{equation}
p = \frac{\partial L}{\partial {\dot \rho}} ,
\label{moment}
\end{equation}

\begin{equation}
E = p\dot \rho - L = \dot \rho\frac{\partial L}{\partial {\dot \rho}} - L ,
\label{Hamilt1}
\end{equation}

\begin{equation}
L = \dot \rho \int E \frac{d{\dot \rho}}{{\dot \rho}^2} = 
\dot \rho \int \frac{\partial E}{\partial {\dot \rho}} \frac{d{\dot \rho}}{\dot \rho} - H
\label{Lagran}
\end{equation}
which give for the conjugate momentum
\begin{equation}
p = \int \frac{\partial E}{\partial{\dot \rho}} \frac{d {\dot \rho}}{\dot \rho} .
\label{pif}
\end{equation}
Substituting for $E$ the Eqn.(\ref{square1}) one gets
\begin{equation}
p = M \log (\dot \rho + \sqrt{{\dot \rho}^2 + 1}) + F(\rho) ,
\label{pi}
\end{equation}

\begin{equation}
L = M (\dot \rho \log(\dot \rho + \sqrt{{\dot \rho}^2 + 1}) - \sqrt{{\dot \rho}^2 + 1}) + 
{\dot \rho} F(\rho) ,
\label{L}
\end{equation}
where $F(\rho)$ is an arbitrary function. The choice of this function does 
not affect the Lagrangian equations of motion, and leads to the canonically 
equivalent systems in the Hamiltonian approach. This choice may become 
important when one explores the rather complex structure of the maximal 
analytical extensions of black hole space-times trying to construct a wave 
function covering the whole manifold. But here we do not need such a 
complication and put $F(\rho) = 0$. From Eqn.(\ref{pi}) we get for $\dot \rho$
\begin{equation}
\dot \rho = \sinh\frac{p}{M} ,
\label{rdot}
\end{equation}
and, after inserting this into Eqn.(\ref{square1}) we obtain the following 
Hamiltonian $H$.
\begin{equation}
H = M \cosh\frac{p}{M} - \frac{G M^2 - e^2}{2\rho} .
\label{H}
\end{equation}
It is convenient to make a canonical transformation to the new variables 
($M$ is the bare mass of the shell)
\begin{equation}
x = M \rho , \qquad \Pi = \frac{1}{M} \rho ,
\label{xPi}
\end{equation}
then
\begin{equation}
H = M \left( \cosh\Pi - \frac{G M^2 - e^2}{2x}\right) .
\label{newH}
\end{equation}

Let us now impose the quantum commutation relation
\begin{equation}
[\Pi, x ] = -i .
\label{commut}
\end{equation}
Then, using the coordinate representation with $\Pi = - i\partial/\partial x$ 
and the relation
\begin{equation}
e^{-i\frac{\partial}{\partial x}} \Psi(x) = \Psi(x-i)
\label{shift}
\end{equation}
we obtain the Schroedinger stationary equation
\begin{equation}
H\Psi(x) = E\Psi(x),
\label{quantum}
\end{equation}

\begin{equation}
\frac{M}{2}\left( (\Psi(x+i) + \Psi(x-i) - \frac{G M^2 - e^2}{x} \Psi(x)\right) = 
m\Psi(x) .
\label{mine}
\end{equation}
Introducing the notation $\epsilon = m/M$ we can rewrite this as 
\begin{equation}
\Psi(x+i) + \Psi(x-i) = \left(2\epsilon + 
\frac{\kappa M^2 - e^2}{x}\right) \Psi(x) .
\label{mefinal}
\end{equation}
Some remarks are in order. First, the Schroedinger equation obtained is not 
the differential, but the equation in finite differences. This is a direct 
consequence of the quantization in proper time which is quite natural in 
the framework of General Relativity. Second, the total mass $m$ is not 
arbitrary anymore but it is an eigenvalue of the Hamiltonian operator, 
subject to the condition that for corresponding eigenfunctions the 
Hamiltonian is Hermitian on the positive semi-axes. 
Third, it should be noted also that the step in our finite difference equation is along the 
imaginary axes, so the solutions should have certain analyticity to be continued to the 
complex plane.

The Hamiltonian 
picture we got is not equivalent to that which could be obtained directly 
from the Einstein-Hilbert action. But our procedure self-consistent. 
We started from equations of motion written already in proper time and 
constructed the new Lagrangian which results in just the equations in proper 
time with no restrictions or/and constraints. We do not expect that 
the resulting mass spectrum will be correct. But our model is extremely 
simple and, as we shall see soon, is exactly solvable, and it seems very 
instructive to develop here some new methods and ideas.

 We consider the non-relativistic limit of the 
obtained Schroedinger equation. With all dimensional quantities restored 
the shifted argument ($x\pm i$) becomes
\begin{equation}
x \pm i \to M \left (\rho \pm \frac{1}{M}i\right) \to 
\rho \pm \frac{\hbar}{Mc}i ,\nonumber
\end{equation}
so for $\rho \gg \hbar/Mc$ we obtain the non-relativistic limit. Expanding 
Eqn.(\ref{mefinal}) up to the second order in ($\hbar /Mc$) we get
\begin{equation}
- \frac{1}{2M}\frac{d^2 \Psi}{d\rho^2} - 
\frac{G M^2 - e^2}{2\rho} \Psi = (E - M)\Psi
\label{nonrel}
\end{equation}
which is just the non-relativistic Schroedinger equation for the radial s-wave 
function, ($E-M$) being the non-relativistic energy of the system. 
For the negative values of ($E-M$) acceptable solutions exist only for a 
discrete spectrum of energies, namely
\begin{equation}
(E- M)_n = - \frac{M(G M^2 - e^2)^2}{8n^2} ,\qquad 
n = 1,2,...
\label{Rydberg}
\end{equation}
For $G = 0$ it reduces to the well known Rydberg formula for the hydrogen atom.

Now, we show how to solve the obtained Schroedinger equation in 
finite differences. And we will do this first in the momentum representation.
 
It is convenient to introduce the following parameters, $\lambda$ and $\beta$,
\begin{equation}
\epsilon = \cos\lambda,\qquad \alpha = G M^2 - e^2 = 2\beta\sin\lambda ,
\nonumber
\end{equation}
in terms of which our Schroedinger equation becomes
\begin{equation}
\Psi(x+i) + \Psi(x-i) = 2\left(\cos\lambda + 
\frac{\beta\sin\lambda}{x}\right) \Psi(x) .
\label{main}
\end{equation}
Since for the black holes $\alpha > 0$, the signs of $\lambda$ and $beta$ are 
the same. 
We choose $\lambda > 0$ and, hence, $\beta > 0$.
Let us return to the operator form of the equation,
\begin{equation}
\left(\cosh\hat p - \beta \sin\lambda \hat x^{-1}\right) \Psi = 
\cos\lambda \Psi .
\label{operatoreqn}
\end{equation}
In the momentum representation operators $\hat p$ and $\hat x$ act as follows
\begin{equation}
\hat p \Psi_p = p \Psi_p ,\qquad \hat x \Psi_p = 
i \frac{\partial}{\partial p} \Psi_p ,
\label{operpx}
\end{equation}
where$|Psi_p$ is a wave function in the momentum representation.
The operator $\hat x^{-1}$ is not well defined. The ambiguity can be removed 
by adding suitable terms to the potential which are proportional to 
$\delta$-function and its derivatives at the origin. But, instead, we can 
multiply the equation by operator $\hat x$ from the left. By doing this we get
\begin{equation}
i\frac{\partial}{\partial p}(\cosh p - \cos\lambda)\Psi_p = 
\beta\sin\lambda \Psi_p ,
\label{dp}
\end{equation}
or
\begin{equation}
\frac{\partial}{\partial p} \log \Psi_p = - \frac{\sinh p + i\beta\sin\lambda}
{\cosh p - \cos\lambda} .
\label{logpsi}
\end{equation}
After introducing a new variable, $z = e^p$, the Eqn.(\ref{logpsi}) takes the 
form
\begin{eqnarray}
\frac{\partial}{\partial z} log \Psi_p = - \frac{z^2 + 
2i\beta\sin\lambda z _ 1}
{z(z^2 - 2\cos\lambda z + 1)} = \frac{1}{z} - \frac{\beta + 1}{z - z_0} + 
\frac{\beta - 1}{z - \bar z_0} ,\nonumber\\
\nonumber\\
\mbox{where} \qquad z_0 = e^{i\lambda},\qquad \bar z_0 = e^{-i\lambda} .
\label{dzlog}
\end{eqnarray}
The above equation can be easily solved, the result is
\begin{eqnarray}
\Psi_p = C \frac{z}{(z - z_0)(z - \bar z_0)} 
\left(\frac{z - \bar z_0}{z - z_0}\right)^{\beta} ,\\
\nonumber\\
z = e^p .\nonumber
\label{solutionp}
\end{eqnarray}
This solution has very important property, it is periodical with the pure 
imaginary period $2\pi i$. This property will be explored below.

Now we will transform the solution found in the momentum representation, to 
the coordinate representation. It can be done by the inverse Fourier transform,
\begin{equation}
\Psi(x) = \frac{1}{\sqrt{2\pi}} \int\limits_{-\infty}^{\infty} 
e^{ipx}\Psi_p dp .
\label{inverse}
\end{equation}
We obtained before the unique (up to the multiplicative constant) solution 
for $\Psi_p$ , namely, Eqn.(\ref{solutionp}). But due to the periodicity 
of $\Psi_p$ we can shift the argument $p\to p+2\pi ki (k=0,\pm 1,\pm 2,...)$  
which will result in the shift of the path of integration in the complex 
momentum plane from the real axis to the parallel one. And after such a shift 
we again obtain a solution in the coordinate representation. But,
\begin{eqnarray}
\Psi_1(x) = \frac{1}{\sqrt{2\pi}}\int\limits_{-\infty}^{\infty} 
e^{ipx}\Psi_p dp ,\nonumber\\
\nonumber\\
\Psi_2(x) = \frac{1}{\sqrt{2\pi}}\int\limits_{-\infty}^{\infty} 
e^{i(p+2\pi ki)x} \Psi_{p+2\pi ki} dp = \nonumber\\
e^{-2\pi kx} \frac{1}{\sqrt{2\pi}}\int\limits_{-\infty}^{\infty} 
e^{ipx}\Psi_p dp,\nonumber\\
\nonumber\\
\Psi_2(x) = e^{-2\pi kx} \Psi_1(x) .
\label{shifts}
\end{eqnarray} 
Thus, given one solution (say, $\Psi_0$), we can construct in this way a 
countable number of solutions solutions, and, in general,
\begin{equation}
\Psi_{general}(x) = \left(\sum_{k=-\infty}^{\infty} c_k e^{-2\pi kx}\right) 
\Psi_0(x) .
\label{fourier}
\end{equation}
The infinite sum in parenthesis is nothing more but a Fourier series for a 
periodical function with an imaginary unit period $i$. And this reproduces 
the following property of the solutions in coordinate representation. Given 
some solution, say, $\Psi_0$ we can construct new solution, $\Psi_1$, 
multiplying it by any periodical function $C(x)$ with the imaginary period 
$i$. 

How many different solutions do we need to construct the general solution? 
The very fact that in the momentum representation we obtained essentially one 
solution proves that for this we need only one solution $\Psi_0$ which 
we call the fundamental solution (though it deserves the name 
superfundamental). Now we will construct the particular fundamental solution 
by the suitable choice of the contour of integration in the 
inverse Fourier transform integral. Note, first of all, that our solution, 
Eqn.(\ref{inverse}), in momentum representation has countable number on 
branching points in the complex momentum plane which can be combined 
in pairs, $(i\lambda+2\pi ki, -i\lambda+2\pi(k+1)i), k=0,\pm 1,\pm 2,...$. 
Connecting two branching points in each pair by a cut we obtain the complex 
plane with countable number of cuts. On the corresponding Riemann surface our 
solution is a single valued analytical function of complex variable. 
Our choice of the contour of integration is as follows. We will integrate 
first along the real axes from left to right (i.e., from 
$-\infty$ to $+\infty$), then along a short curve at the right infinity 
$(p\to p+2\pi i, p\to{+\infty})$, then along the straight line $y=2\pi i$, 
parallel to the real axis, from left to right, and finally along a short curve 
at the left infinity back to the real axis. The integration along the short 
curves at infinities between the straight lines gives zero contributions for 
positive values of $x$ (for negative $x$ we can choose the straight line 
$y=-2\pi i$ instead of $y=2\pi i$). The integration along each straight line 
gives us a solution the linear combination of which is again a solution. 
Thus, the inverse Fourier integral along such a closed contour gives us 
a solution. This contour can be distorted to become a contour around the cut 
$(\lambda i, (2\pi-\lambda)i)$  for $x>0$ (or around the cut 
$(-\lambda i, -(2\pi -\lambda)i)$ for $x<0$). Thus,
\begin{equation}
\Psi_0(x) = \frac{1}{\sqrt{2\pi}}\oint\limits_{C_+} e^{ipx} \Psi_p dp , 
\qquad x > 0 , 
\label{psipos}
\end{equation}  

\begin{equation}
\Psi_0(x) = \frac{1}{\sqrt{2\pi}}\oint\limits_{C_-} e^{ipx} \Psi_p dp , 
\qquad x < 0 , 
\label{psineg}
\end{equation}
where
\begin{eqnarray}
\Psi_p = C \frac{z}{(z - z_0)(z - \bar z_0)} \left(\frac{z - 
\bar z_0}{z - z_0}
\right)^{\beta}\\
\nonumber\\
z = e^p,\qquad z_0 = e^{i\lambda},\qquad \bar z_0 = e^{-i\lambda}\nonumber
\label{psimoment}
\end{eqnarray}
In what follows we restrict ourselves to the case of positive $x$ only.
The above integral representation for $\Psi_0$ can be simplified if we 
integrate Eqn.(\ref{psipos}) by parts. Using the fact that
\begin{equation}
\frac{zdp}{(z-z_0)(z-\bar z_0)} = - \frac{1}{2i\sin\lambda} 
d\left(\log{\frac{z-\bar z_0}{z-z_0}}\right) ,
\end{equation}
we get
\begin{equation}
\Psi_0(x) = x \oint\limits_{C_+} e^{ipx} \left(\frac{z-\bar z_0}
{z-z_0}\right)^{\beta} dp ,\qquad x> 0 ,
\label{intrepr}
\end{equation}
(the extra term vanishes because our contour has no boundary and for 
convenience we omitted some constant factors). We will use use 
Eqn.(\ref{intrepr}) as an integral representation for the fundamental 
solution of our finite differences equation, Eqn.(\ref{main}).

Let us investigate the fundamental solution in 
more details. Our aim is to reduce the integration along the closed 
contour $C_+$ around the cut to the finite interval between the 
corresponding branching points. But for $\beta\ge 1$ (remember that we 
have chosen $\beta$ positive) we have the non-integrable singularity at 
the lower integration limit. To avoid this difficulty we need some 
recurrent relation for lowering the parameter $\beta$. To find such a 
relation we integrate by parts Eqn.(\ref{intrepr}), the result is the 
following.
\begin{equation}
\Psi_{\beta}(x) = \Psi_{\beta -1}(x) + \frac{ix}{\beta -1}\left\{
\Psi_{beta -1}(x) - e^{-i\lambda} \Psi_{\beta -1}(x+i)\right\} .
\label{recurrent}
\end{equation}
From the above relation it is easy to derive the structure of 
$\Psi_{\beta}$ for general values of $\beta >1$. Let us take 
$\beta =n+\tilde\beta$, with $\tilde\beta \le 1$. Then, $\Psi_{\beta}$ is 
the sum of two terms, which of them is the product of some polynomial of 
$n$-th degree and of $\Psi_{\tilde{\beta}}(x)$ or $\Psi_{\tilde{\beta}}(x+i)$. 
Thus, we can proceed assuming $\beta\le 1$.
First of all, consider separately the case $\beta =1$. We have
\begin{eqnarray}
\Psi_1 = x \oint\limits_{C_+} e^{ipx}\frac{e^p - e^{-i\lambda}}
{e^p - e^{i\lambda}} dp &=& 
2\pi ix \lim_{p\to i\lambda} \frac{e^{ipx}(e^p - e^{-i\lambda})
(p - i\lambda)}{e^p - e^{i\lambda}} \nonumber\\
&=& - \left(4\pi e^{-i\lambda}\sin\lambda\right) xe^{-\lambda x} .
\label{beta1}
\end{eqnarray}
From this it follows that all $\Psi_{\beta}$ for positive integer 
$\beta =n$ has the form
\begin{equation}
\Psi_n = P_n(x) e^{-\lambda x} ,
\label{psin}
\end{equation} 
where $P_n(x)$ are some polynomials of $n$-th degree.
In the case $0<\beta <1$ the integrand in the Eqn.(\ref{intrepr}) is 
integrable on both ends of the cut in the complex momentum plane and we 
are able to convert the integral along the closed contour $C_+$ into the 
integral along the finite interval. We get the following,
\begin{equation}
\Psi_{\beta}(x) = \oint\limits_{C_+} e^{ipx} f_p dp = \left(1 - 
e^{2\pi i\beta}\right) x \Phi_{\beta}(x) ,
\end{equation}
where
\begin{equation}
\Phi_{\beta}(x) = \int\limits_{i\lambda}^{i(2\pi -\lambda)} e^{ipx}
\left(\frac{e^p - e^{-i\lambda}}{e^p - e^{i\lambda}}\right)^{\beta} dp ,
\qquad x > 0 .
\label{phi}
\end{equation}
Changing the variables,
\begin{equation}
e^q - e^{i\lambda} = -(2i\sin{\lambda}) y ,
\label{qy}
\end{equation}
we get
\begin{equation}
\Phi_{\beta} = \left(1 - e^{-2i\lambda} \right) e^{-2\lambda x} 
\int\limits_0^1 
\left(1 - \left(1-e^{-2i\lambda}\right) y \right)^{ix-1} \left(
\frac{y - 1}{y}\right)^{\beta} dy .
\label{phibeta}
\end{equation}
Comparing this with the well known integral representation for the Gauss's 
hypergeometric function, $F(a,b;c;z)$,
\begin{eqnarray}
F(a,b;c;z) = \frac{1}{B(b,c-b)} \int\limits_0^1 t^{b-1} (1 - t)^{c-b-1} 
(1 - tz)^{-a} dt ,
\nonumber\\
B(x,y) = \int\limits_0^1 t^{x-1} (1 - t)^{y-1} dt ,
\label{Gauss}
\end{eqnarray}
we see that the integral in Eqn.(\ref{phibeta}) is the Gauss's hypergeometric 
function $F(a,b;c;z)$ with the following values of parameters,
\begin{equation}
a = 1 - ix,\qquad b = 1 - \beta;\qquad c = 2;\qquad z = 1 - e^{-2i\lambda} , 
\label{param}
\end{equation}
and
\begin{equation}
B(b,c-b) = B(1-\beta,1+\beta) = \frac{\pi \beta}{\sin{\pi \beta}} .
\label{B}
\end{equation}
Finally, for the fundamental solution $\Psi_{\beta}(x)$ we get
\begin{equation}
\Psi_{\beta}(x) = \left(-4\pi\beta e^{-i\lambda} \sin\lambda\right) 
x e^{-\lambda x} 
F\left(1 - ix, 1 - \beta; 2; 1 - e^{-2i\lambda}\right) .
\label{fundhyper}
\end{equation}
The above expression was derived for $0 < \beta , 1$ only, but the same is 
valid for any value $\beta > 0$.
And, finally, it can be easily shown that the original equation in 
finite differences, Eqn.(\ref{main}), is a direct consequence of the 
following Gauss's recurrent relation 
\begin{equation}
(2a-c-az+bz) F(a,b;c;z) + (c-a) F(a-1,b;c;z) + a(z-1) F(a+1,b;c;z) = 0
\nonumber
\end{equation}

So, we found the general solution to the equation in 
finite differences with Coulomb potential,
\begin{equation}
\Psi(x + i) + \Psi(x - i ) = 2 \left(\cos{\lambda} + \frac{\beta}{x} 
\sin{\lambda}\right) \Psi(x) .
\label{ours}
\end{equation} 
Now we should remember that this equation is actually the radial 
Schroedinger equation for zero angular momentum, and the coordinate $x$ runs 
from zero to infinity, i.e., $x$ takes values in the positive semi-axis only. 
Thus the Hamiltonian,
\begin{equation}
\hat H = \cosh{\left( i\frac{\partial}{\partial x}\right)} - 
\frac{\beta}{2x} \sin{\lambda}
\label{Hamilt}
\end{equation}
should be the selfadjoint operator on the positive semi-axis rather than on 
the whole real axis as dictated by the quantum mechanics postulates. 
In addition, the wave function should be square integrable on the semi-axis.
The corresponding extension of the above Hamiltonian was found by Petr 
Hajicek. 
It appeared the wave function should obey the following boundary conditions 
at the origin, $x = 0$ ,
\begin{equation}
\Psi^{(2n)}(0) = 0, \qquad  n = 0,1,...  .
\label{boundorigin}
\end{equation}
That is, the function itself and all its even derivatives should be zero at 
the origin. The appearance of infinite number of conditions is due to the 
infinite order of the operator. P.Hajicek found also a conserved probability 
current $J(x)$ for the Schroedinger equation in finite differences, 
Eqn.(\ref{ours}) . It is given as usual by the equation
\begin{equation}
J^{\prime}(x) = i \left(\Psi^* {\hat H} \Psi - \Psi {\hat H} \Psi^* \right)\nonumber
\end{equation} 
so that 
\begin{equation}
\left( \Psi^{*} \Psi \right)^. + J^{\prime} = 0\nonumber
\end{equation}
Writing the Hamiltonian (\ref{Hamilt}) as follows
\begin{equation}
\hat H = \frac{1}{2} \sum_{k=0}^{\infty} \frac{(-1)^k}{(2k)!} \Psi^{(2k)}(x) - 
\frac{\beta\sin{\lambda}}{2x}
\label{Hamiltser}
\end{equation}   
we obtain for $J(x)$ ,
\begin{equation}
J(x) = i \sum_{k=0}^{\infty} \frac{(-1)^k}{(2k)!} \sum_{l=1}^k (-1)^{l-1}
\left[\Psi^{*(l-1)} \Psi{(2k-l)} - \Psi^{(l-1)} \Psi^{*(2k-l)}\right] .
\label{currenthaj}
\end{equation} 
P.Hajicek showed that the boundary condition (\ref{boundorigin}) implies that
\begin{equation}
J(0) = 0 .
\label{currentzero}
\end{equation}
The boundary condition (\ref{boundorigin}) is a direct generalization of 
the boundary condition in the case of non-relativistic Schroedinger equation 
which is of second order. In this particular case the boundary conditions 
(\ref{boundorigin}) are reduced to the single one, $\Psi(0)=0$ , which is 
together with the square integrability condition ensures existence of a 
discrete spectrum for bound states. And, in general, if we expand 
Hamiltonian (\ref{Hamilt}) in series of derivatives and consider  
the truncated Hamiltonian of $2N$th order, the corresponding Schroedinger 
equation becomes the differential equation of $2N$th order. It can be shown 
that the asymptotics at of such an equation at the origin are
\begin{equation}
\Psi \sim x , x^2 , ... , x^{2N-1} , 1 + x^{2N-1} \log x 
\label{asymptrunc}
\end{equation}
and at the infinity they are of the form
\begin{equation}
\Psi \sim e^{\pm \lambda_k x} e^{\alpha_k\log x} ,\qquad k=1,...N .
\label{asyptrunc2}
\end{equation}
where $\lambda_k$ are real for bound states. The general solution is defined 
up to a normalization factor and depends, actually, on $(2N-1)$ arbitrary 
constants which are to be determined using boundary conditions. The square 
integrability condition reduces the number of unknown constants to $(N-1)$. 
But for the truncated operator of $2N$th order we have $N$ conditions at 
the origin. They can only be satisfied for a discrete spectrum 
of eigenvalues. Note that the last of the asymptotics (\ref{asymptrunc}) is 
excluded by the boundary condition at the origin. We can reverse the above 
consideration and start from the asymptotic behavior at the origin. 
To satisfy the boundary conditions there we have to exclude the last 
asymptotics (\ref{asymptrunc}) and all the even terms in the expansion up 
to the $(2N-2)$th order. By doing this can determine $N$ of $(2N-1)$ unknown 
constants. Thus we are left with only $(N-1)$ constants for N boundary 
conditions at infinity. But the situation is not so simple in the case of an 
infinite order operator. First of all, how to get rid of the logarithmic 
term in asymptotics? The problem is that if we first go to the infinite 
order limit, then the logarithmic term unconditionally disappears 
$(|x|<1 !)$ leaving $1$ as an asymptotics. But the latter can be easily 
reproduced by some infinite linear superposition of the remained asymptotics. 
Subtracting one from another we would obtain the solution which would satisfy 
formally all the boundary conditions at the origin still having enough 
arbitrary constants to satisfy boundary conditions at infinity.
But if we first differentiate $2N$ times the asymptotic solution containing 
logarithmic term we will get ($1/x$) term. Thus, requiring that the ``good'' 
solution should be infinitely differentiable we can reach our goal. 
Moreover, we have to require the analyticity of the wave function at least on 
the real axis. Thus is because we implicitly have used the analyticity of 
the solutions in transition from the differential equation of infinite order 
to the finite differences equation. The situation is not good at the infinity 
either. We saw that in the case of our particular equation we can obtain a 
new solution multiplying the known one by $\exp{(-2\pi kx)}$ . 
Choosing the large enough value of $k$ we are able to convert the ``bad'' 
at infinity solution to the ``good'' one. We will see in a moment that it is 
the analyticity condition which solves this problem.

Let us consider the asymptotics of the general solution at 
$x \to 0$ and $x \to \infty$ . The idea is the following. First, we find the
asymptotical behavior of the fundamental solution at $x \to 0$ and
 $x \to \infty$. We have,
\begin{equation}
\label{aaz}
x^r,\qquad r = 1,2... .
\end{equation}
at $x \to 0$ and 
\begin{eqnarray}
\Psi_{\beta} &=& - 2\pi i \beta e^{-i\lambda\beta} e^{i\pi\beta} \left\{
\frac{(2\sin{\lambda})^{\beta}}{\Gamma(1+\beta)} x^{\beta} e^{-\lambda x} -
\frac{(2\sin{\lambda})^{-\beta}}{\Gamma(1-\beta)} x^{-\beta} e^{\lambda x}\right\}
\phi(\frac{1}{x}) ~,
\nonumber \\
 x &\to& \infty 
\label{inftyasymp}
\end{eqnarray}
It is interesting to note that the same asymptotics can be found directly 
from the original Schroedinger equation in finite differences without 
knowing the exact solution.

Comparing the asymptotics at $x = 0$ and at infinity we see that the 
fundamental solution is an analytic function only if
\begin{equation}
\beta = n
\label{betan}
\end{equation}
(remember that we chose $\beta >0$).
This leads to the discrete spectrum for the total mass $m$. Indeed,we have
\begin{eqnarray}
\beta = \frac{\alpha}{2\sin{\lambda}} = n ,\qquad \alpha = \kappa M^2 ,\nonumber\\
\epsilon = \frac{m}{M} = \cos{\lambda} ,\nonumber\\
\epsilon = \sqrt{1 - \frac{\alpha^2}{4n^2}}
\label{epsilonn}
\end{eqnarray}
For $\beta=n$ the hypergeometric series in Eqn(\ref{fundhyper}) terminates and we are left 
with the following fundamental solution,
\begin{equation}
\Psi_n(x) = P_n(x) e^{-\lambda x} ,
\label{solpoly}
\end{equation}
where $P_n$'s are some polynomials of order $n$.
They can be calculated directly from the 
definition, or from the generating function
\begin{eqnarray}
\Phi(z, x) = \sum_{n=0}^{\infty} \frac{\Phi^{(n)}(0, x)}{n!} z^n = 
\sum_{n=0}^{\infty} \Pi_n(x) z^n = \nonumber\\
2\pi e^{2\lambda x} \left( \frac{z + \cot{\lambda} + i}{z +\cot{\lambda} - i}\right)^{ix}
\label{genfunc}
\end{eqnarray}

At the end of this section we give an example of solving the boundary 
conditions and find the wave function of the ground state (i.e., for $n=1$).
The fundamental solution for $n=1$ was found to be 
\begin{equation}
\Psi_1 = x e^{-\lambda x}
\label{ps1}
\end{equation}
(we omitted here the irrelevant constant factor). The general solution 
obeying the boundary condition at infinity (exponential falloff at 
$x\to\infty$) has the form
\begin{equation}
\Psi_{1 gen} = x e^{-\lambda x} \sum_{k=0}^{\infty} c_k e^{-2\pi kx} ,
\label{ps1gen}
\end{equation}
and the coefficients $c_k$ are to be determined  by solving boundary 
conditions at the origin,
\begin{equation}
\Psi^{(2l)}(0) = 0,\qquad l = 0, 1, ... 
\label{orig}
\end{equation}
Differentiating Eqn.(\ref{ps1gen}) $2l$ times we get the following infinite 
set of linear equations, 
\begin{equation}
2l\sum_{k=0}^{\infty} c_k (\lambda + 2\pi k)^{2l-1} = 0, \qquad l=0,1,...
\label{linset}
\end{equation}
The determinant of this system is identically zero because the first line consists of 
zeros. We can truncate this system at some specific value of $l$, calculate 
all the determinants and minors and then, after appropriate renormalization 
(say, putting $c_0=1$), take the limit $l\to\infty$. In our simplest case of 
the ground state it is possible to make all the calculations up to the very 
end with the following result
\begin{equation}
c_k = \frac{(-1)^k}{k!} \frac{\Gamma(\frac{\lambda}{\pi} + k)}
{\Gamma(\frac{\lambda}{\pi})} . \qquad k=0,1,...
\label{ck}
\end{equation}  
Inserting this into Eqn.(\ref{ps1gen}) we can write the ground state wave 
function (up to the normalization factor) in a very simple and nice form,
\begin{eqnarray}
\Psi_1 = x e^{-\lambda x} \sum_{k=0}^{\infty} c_k e^{-2\pi kx} =\nonumber\\
x e^{-\lambda x} \sum_{k=0}^{\infty} \frac{(-1)^k}{k!}\frac{\lambda}{\pi}
\left(\frac{\lambda}{\pi}+1\right)\ldots 
\left(\frac{\lambda}{\pi}+k-1\right)e^{-2\pi kx} =\nonumber\\
x e^{-\lambda x}\left(1 + e^{-2\pi x}\right)^{-\frac{\lambda}{\pi}} =
\nonumber\\
2^{-\frac{\lambda}{\pi}}\frac{x}{(\cosh{\pi x})^{\frac{\lambda}{\pi}}} .
\label{wavegr}
\end{eqnarray}
It is easy to see that after shifting the parameter 
$\lambda\to\lambda+2\pi l$, where $l$ is a positive integer, we again obtain 
a solution satisfying all the boundary conditions (technically such a 
solution can be obtained if put zero first $l$ coefficients 
$c_m, m=0,...(l-1)$ in the infinite set of linear equations considered 
above and normalize to unity $c_l$). Moreover, the introduced earlier 
conserved current is identically zero for any superposition (with complex 
coefficients) of these functions. This means that each eigenvalue of a total 
mass (energy) is infinitely degenerate and the appearance of the infinite number
of branching point reflects this fact.

\subsection{Quantum black holes and Hawking's radiation.}

In the preceding sections we have got a complete solution to the Schroedinger 
equation in finite differences describing a quantum mechanical behavior of 
the self-gravitating electrically charged spherically symmetric dust shell. 
For the bound states we found the discrete spectrum of mass which coincides 
with Dirac spectrum if we put zero the so called the so called radial quantum 
number. Similarly to the case of hydrogen atom, to which our spectrum reduces 
in the non-relativistic limit, we may conclude that the shell does not 
collapse without radiation. The starting classical situation is however 
different. The classical hydrogen atom collapses radiating continuously while 
the classical spherically symmetric shell collapses without radiation. 
But without such a collapse the event horizon and thus the black hole can not 
be formed at all. Moreover, if we calculate the mean value of radius of the 
shell using the wave function we will find that at least for large values of 
principal quantum number it is far away of classical event horizon, so the 
shell spends most of its ``lifetime'' outside the classical black hole. 
So, to get a black hole solution we need some new criterion and more detailed 
investigation of the spectrum is needed. Let us consider the dependence of a 
total mass $m$ on a bare mass $M$ , the other variables, $e$ and $n$ fixed. 
We see that there are two branches, an increasing one for
\begin{equation}
(G M^2 - e^2)(3 G M^2 - e^2)<4n^2
\label{branch}
\end{equation}
and a decreasing one in the opposite case. Note that now
\begin{equation}
\frac{m}{M} > \sqrt{\frac{2}{3}} \sqrt{1+ \frac{e^2}{3 G M^2 - e^2}} ,
\label{nmineq}
\end{equation}
and this value is greater than the corresponding classical value.
The increasing branch corresponds to the ``black hole case'' while the 
decreasing branch - to the ``wormhole case''. Using a ``quasiclassical'' 
argumentation we can say that for the states obeying inequality 
(\ref{branch}) the expectation value of the radius lies outside the event 
horizon on the ``our'' side of the Einstein-Rosen bridge (thus replacing 
the notion of ``classical turning point'' by that of ``radius expectation 
value''). For the decreasing branch the same occurs on the ``other'' side of 
the Einstein-Rosen bridge. It is clear now why the value of total mass-bare 
mass ratio in quantum case is greater than the corresponding classical value. 
The origin of this is just the replacing of the classical turning point by 
the mean value which is smaller thus giving rise to the increase in the 
total mass. Following such a line of reasoning we should conjecture 
that the maximal possible value of a total mass $m$ for charge $e$ and 
principal quantum number $n$ fixed corresponds to the situation when the 
mean value of the radius of the shell lies at the event horizon 
making its collapse possible. The further increase in the bare mass $m$ will 
lead to the wormhole with smaller total mass. Generalizing, we can introduce 
the notion of the quantum black hole states. For given values of electrical 
charge $e$ and principal quantum number $n$ the quantum black hole state is 
the state with maximal possible total mass $m$. Thus, for the quantum black 
hole states we have, instead of inequality (\ref{branch}) the following 
equation
\begin{equation}
3(G M^2)^2 - 4 G M^2e^2 + e^4 - 4n^2 = 0 ,
\label{brancheq}
\end{equation} 
the solution of which subject to the condition $G M^2-e^2>0$ is
\begin{equation}
G M^2 = \frac{2}{3}e^2 + \frac{1}{3}\sqrt{e^4 + 12n^2} .
\label{root}
\end{equation}
For the black hole mass spectrum we get eventually
\begin{eqnarray}
m_{BH} = \frac{1}{\sqrt{G}} \sqrt{\frac{4\sqrt{3}}{9}n \left(1 + 
\frac{e^4}{12n^2}\right)^{3/2} + \frac{2}{3}e^2 - \frac{e^6}{54n^2}}\\
n = 1, 2, 3,...\nonumber
\label{BH}
\end{eqnarray}
For the uncharged black hole with $e=0$ we have immediately
\begin{equation}
m = \frac{2}{\stackrel{4}{\sqrt{27}}}\sqrt{n}m_{Pl} .
\label{BHm}
\end{equation}
Two limiting cases are of interest. In the first case
\begin{eqnarray}
\frac{e^4}{12n^2} \ll 1,\nonumber\\
\nonumber\\
m_{BH} = \frac{2}{\stackrel{4}{\sqrt{27}}}\frac{\sqrt{n}}{\sqrt{G}}
\left(1 + \frac{\sqrt{3}}{4}\frac{e^2}{n} - \frac{1}{32}\frac{e^4}{n^2} + 
\ldots\right) ,
\label{one}
\end{eqnarray}
and the charge of the black hole gives rise only to the small corrections 
to the mass spectrum of the Schwarzschild black hole. The corresponding 
expressions for the bare mass $M$ and the ratio $m/M$ are, respectively,
\begin{eqnarray}
M = \stackrel{4}{\sqrt{\frac{4}{3}}}\frac{\sqrt{n}}{\sqrt{G}}
\left(1 + \frac{\sqrt{3}}{6}\frac{e^2}{n} - \frac{1}{48}\frac{e^4}{n^2} + 
\ldots\right)\nonumber\\
\nonumber\\
\frac{m}{M} = \sqrt{\frac{2}{3}}
\left(1 + \frac{\sqrt{3}}{12}\frac{e^2}{n} - \frac{5}{96}\frac{e^4}{n^2} + 
\ldots\right) .
\label{MmM}
\end{eqnarray}
In the second case
\begin{eqnarray}
\frac{e^4}{12n^2} \gg 1,\nonumber\\
\nonumber\\
m_{BH} = \frac{|e|}{\sqrt{G}}\left(1 + \frac{n^2}{2e^4} + \ldots\right) .
\label{extr}
\end{eqnarray}
The corresponding expressions for $M$ and $m/M$ are, respectively,
\begin{eqnarray}
M = \frac{|e|}{\sqrt{G}}\left(1 + \frac{n^2}{e^4} + \ldots\right) ,\nonumber\\
\nonumber\\
\frac{m}{M} = 1 - \frac{n^2}{2e^4} + \ldots .
\label{MmM2}
\end{eqnarray}
It is interesting that in this second limiting case the black hole mass takes 
values nearly the mass of a classical extreme Reissner-Nordstrom black hole.
It is easily seen that in both limiting cases $|\bigtriangleup m|/m\ll 1$, 
where $|\bigtriangleup m|$ is the difference in masses for nearby energy 
(mass) levels. So, these limits are essentially quasiclassical ones, and the 
corresponding mass spectra should be compatible with the existence of 
Hawking's radiation. This is indeed the case. Let us write once more the 
spectra that follow from the black hole thermodynamics in the same 
limiting cases.
\begin{eqnarray}
m &=& \frac{\sqrt{n}}{2\sqrt{G}}\sqrt{\alpha_1} \left(1 + 
\frac{\beta_1+3}{2\alpha_1} \frac{e^2}{n} \right.
\nonumber \\
&+& 
\left. \left(\frac{\gamma_1}{2\alpha_1} - 
\frac{(\beta_1+1)(\beta_1+4)}{8\alpha_1^2}\right)\frac{e^4}{n^2} +
\ldots\right) ,\\
\nonumber\\
&&n \gg e^2\nonumber
\end{eqnarray}
and
\begin{eqnarray}
m = \frac{|e|}{2\sqrt{G}} \left(2 + \frac{\alpha_2^2}{4} \frac{n^2}{e^4} +
\ldots\right) ,\\
\nonumber\\
&&n \ll e^2\nonumber
\end{eqnarray}

Comparing this with the expansion obtained from the black hole spectrum, 
we have
\begin{eqnarray}
\alpha_1 = \frac{16}{3\sqrt{3}},\qquad \beta_1 &=& -\frac{1}{3},
\qquad \gamma_1 = \frac{5\sqrt{3}}{144},...\nonumber\\
\alpha_2 = 2,\qquad \beta_2 &=& 2,\qquad \gamma_2 = -15,... 
\end{eqnarray}
Comparing values for $\alpha_1$ and $\alpha_2$ we see that $\tilde\alpha$ is 
indeed a slowly varying function. Thus, we see that in the limiting case of 
low temperature our black hole spectrum is compatible with the existence of 
Hawking's radiation. Finally, we would like to note that our black hole 
spectrum obeys also the third law of thermodynamics. Indeed, for a nearly 
extreme Reissner-Nordstrom black hole the lowest energy level (for $n=1$) 
always exceeds the critical value $(=|e|/\sqrt{\kappa})$, so the zero 
temperature state is not accessible.

\section{Geometrodynamics for Black Holes and Wormholes.}

This Chapter is devoted to the ``correct'' quantization of the spherically 
symmetric gravity with thin shells. It means that we start with the 
Einstein-Hilbert action reduced to the spherically symmetric space-time, 
make use of the Arnowitt-Deser-Misner (ADM) formalism in order to obtain 
the canonical constraints and Hamilton equations of motion and formulate 
the quantum theory. 

Our model remains the same. This is just a self-gravitating spherically 
symmetric dust thin shell, endowed with a bare mass $M$. The whole 
space-time is divided into three different regions: the inner part ( I ), 
the outer part ( II ) containing no matter fields separated by thin 
layer III, containing the dust matter of the shell.
The general metric of a spherically symmetric space-time is now 
convenient to write in the form:
\begin{equation}
\label{metric}
ds^2=- N^2 dt^2 + L^2 ( dr + N^r dt )^2 + R^2 ( d\theta^2 + \sin^2\theta
d\phi^2 )
\end{equation}
where $(t,r,\theta ,\phi )$ are space-time coordinates, $ N, N^r, L, R$ are
some functions of $t$ and $r$ only. Trajectory of the thin shell is some
3-dimensional surface $\Sigma $ in space-time  given by
some function $\hat r(t)$: $\Sigma^3=\{ (t,r,\theta ,\phi ): r=\hat r(t)\} $. 
In region I $ r< \hat r-\epsilon $,
in a region II $ r>\hat r+\epsilon $, region III is a thin layer 
$\hat r-\epsilon <r <\hat r+\epsilon $.
We require that metric coefficients $N,N^r,L$ and $R$ are
continuous functions but jump discontinuities could appear in their 
derivatives at the points of $\Sigma $ when the limit $\epsilon\rightarrow 0$ 
is taken.
The action functional for the system of spherically symmetric gravitational 
field and the thin shell is
\begin{equation}
\label{action}
\begin{array}{c}
S=S_{gr}+S_{shell}=\frac{1}{16\pi G}\int\limits_{I+II+III}\
^{(4)}R\sqrt{-g} d^4x + (\mbox{surface terms}) - \\ - M\int\limits_{ \Sigma } 
d\tau
 \end{array}
 \end{equation}
It consists of the standard Einstein-Hilbert action for the gravitational 
field and matter part of the action describes a thin shell of dust. 
The surface terms in the gravitational action and the falloff behavior of 
the metric and
its derivatives were studied in details by Karel Kuchar. So we will not 
consider this question and will use the results of Kuchar when needed. We will 
be interested in the behavior of the action and constraints on the 
surface $\Sigma^3$ representing the shell's trajectory.   
The complete set of degrees of freedom of our system consists of the set of
$N(r,t), N^r(r,t), L(r,t), R(r,t)$ which describe gravitational field
and $\hat r(t)$ which describes the motion of the shell.
The metric (\ref{metric}) has the standard ADM form for 3+1
decomposition of a space-time with lapse function $N$ , shift vector 
$N^i=(N^r,0,0)$ and space
metric $h_{ik}=\mbox{diag}( L^2, R^2, R^2\sin^2\theta)$ given foliation of the 
manifold on space and time. The scalar curvature density has the form
\begin{equation}
\label{curvature}
\begin{array}{c}
\ ^{(4)}R\sqrt{-g}=\\
N\sqrt{h}\left( \ ^{(3)}R+\left( (\mbox{Tr} K)^2-\mbox{Tr} K^2\right)
\right)- \\
- 2\left( \sqrt{h} K)_{,0} \right) + 2 \left( \sqrt{h} KN^i-\sqrt{h}
h^{ij} N_{,j}\right)_{,i}
\end{array}
\end{equation}
where $\ ^{(3)}R$ and $K^{ij}$ are the  scalar curvature of a space metric 
$h_{ij}$ and exterior curvature tensor of a  surface $t=\mbox{const}$. 
Substituting
expression (\ref{metric}) for the metric into (\ref{curvature}) we obtain
the expression for internal and external curvatures of the surface $t=const$
in the form
\begin{equation}
\ ^{(3)}R=\frac{
\displaystyle
2}{\displaystyle R^2}\left( 1- \frac{\displaystyle (R')^2}{\displaystyle
L^2}-\frac{\displaystyle 2RR''}{\displaystyle L^2}+\frac{\displaystyle
2RR'L'}{\displaystyle L^3} \right)
\end{equation}
and
\begin{equation}
\begin{array}{c}
K^i_j=\mbox{diag}(K^r_r, K^\theta_\theta, K^\phi_\phi ) \\
K^r_r=\frac{\displaystyle 1}{\displaystyle NL}\left( \dot L-L'N^r-L(N^r)'
\right), \\
K^\theta_\theta =K^\phi_\phi =\frac{\displaystyle 1}{\displaystyle NR} \left(
\dot R-R'N^r\right)
 \end{array}
 \end{equation}
Here dot and prime denote differentiation in  $t$ and  $r$ respectively.
Contributions to the gravitational action from the terms containing total
derivatives in (\ref{curvature})  give rise to the surface terms which 
cancel each other at the common boundaries
of regions I, II and II, III. So we are left with the surface terms at 
infinity. We will turn to them later.
The essential part of the action for gravitational
field is just the ADM part of the action (\ref{action}) with Lagrangian 
\begin{equation}
\label{ADM}
L_{gr}=
\frac{1}{16\pi G}NLR^2\left(\ {(3)}R-(\mbox{Tr} K)^2-\mbox{Tr} K^2\right)
\end{equation}
Contribution to the  action from the integral over the region III
in the limit $\epsilon\rightarrow 0$ is only due to the term containing 
second derivative of $R$, namely 
\begin{equation}
\int\limits_{III} \frac{1}{16\pi G}NLR^2\ ^{(3)}R =
-\int\limits_{II} \frac{\displaystyle
NRR''}{\displaystyle GL}=
-\int\limits_{\Sigma } \frac{\displaystyle \hat N
\hat R\left[ R'\right] }{
\displaystyle G\hat L}
\end{equation}
We will denote by hats variables on $\Sigma $ and by 
$\left[ {\cal A}\right] =\lim_{\epsilon\to 0}\left( 
{\cal A}(\hat r+\epsilon)-{\cal
A}(\hat r-\epsilon)\right)$ a jump of variable ${\cal A}(r)$ on the
shell surface.
Substituting the expression (\ref{metric}) into the shell part of
the action we have:
\begin{equation}
\label{matter}
S_{shell}
=
-M\int\limits_\Sigma
\sqrt{ \hat {N}^2-\hat {L}^2 \left(
\hat {N}^r + \dot{\hat {r}}
\right)^2} dt
\end{equation}
The explicit form of the action (\ref{action}) with metric (\ref{metric})
becomes
\begin{eqnarray}
\label{actionone}
 S&=&
\frac{1}{G}\int\limits_{I+II+III}
\left( N
\frac{\displaystyle L}{\displaystyle 2}
\frac{\displaystyle (R')^2}{\displaystyle 2L}-
\left(\frac{\displaystyle RR'}{\displaystyle L}\right) '
+\frac{\displaystyle R}{\displaystyle N}
\left(\dot{R}-R'N^r\right)
\left( (LN^r)'-\dot{L}\right) \right.
\nonumber \\
&+&
\left.
\frac{\displaystyle L}{\displaystyle 2N}
\left( \dot{R}-R'N^r\right)^2
\right)
-\int\limits_{\Sigma }
\left(
\frac{\displaystyle \hat{N}\hat{R}\left[ R'\right] }{\displaystyle \hat{L}}
-m\sqrt{ \hat{N}^2-\hat{L}^2\left( \hat{N}^r+\dot{\hat{r}}\right)^2}
\right) dt
\end{eqnarray}
The canonical formalism for this action can be described in the following 
way. Momenta conjugate to corresponding dynamical variables are
\begin{equation}
\label{momenta}
\begin{array}{rcl}
P_N&=&\delta S\left.\right/\delta \dot N=0;\\
P_{N^r}&=&\delta S\left.\right/\delta \dot N^r=0\\
P_L&=&\delta S\left.\right/\delta \dot L =\frac{\displaystyle R}
{\displaystyle GN}\left(R'N^r-\dot R\right)\\
P_R&=&\delta S\left.\right/\delta \dot R =\frac{\displaystyle L}
{\displaystyle GN}\left( R'N^r-\dot R\right)+
\frac{\displaystyle R}{\displaystyle GN}\left( (LN^r)'-\dot L\right)\\
P_{\hat R}&=&\delta S\left.\right/\displaystyle\delta \dot{ \hat R}=0\\
P_{\hat L}&=&\delta S\left.\right/\delta \dot{\hat L}=0\\
\hat\pi&=&\delta S\left.\right/\delta \dot {\hat r}=\frac
{\displaystyle m\hat L^2 ( N^r+\dot{\hat{r}} ) }
{\displaystyle
\sqrt{\hat N^2-\hat L^2 (N^r+\dot{\hat {r}} ) } }
\end{array}
\end{equation}
The action (\ref{actionone}) rewritten in the Hamiltonian form becomes
\begin{equation}
\label{hamil}
\begin{array}{c}
S=\int\limits_{I+II}\left( P_L\dot L+P_R\dot R-NH-N^rH_r\right) dr dt+
\int\limits_\Sigma \hat{\pi}\dot{\hat{ r}}- \\
\hat N
\left(
\hat R\left[ R'\right] /(G\hat L) + \sqrt{m^2+ \hat\pi^2/ \hat L^2}
\right) -
\\
\hat N^r \left(-\hat L\left[ P_L\right]-\hat\pi\right) dt
\end{array}
\end{equation}
with
\begin{equation}
\label{constraints}
\begin{array}{rcl}
H&=&
G\left( \frac{\displaystyle LP_L^2}{\displaystyle 2R^2}-
\frac{\displaystyle P_LP_R}{\displaystyle R}\right)
+\frac{\displaystyle 1}{\displaystyle G}\left( - \frac{\displaystyle
L}{\displaystyle 2}- \frac{\displaystyle (R')^2}{\displaystyle 2L}+ \left(
\frac{\displaystyle RR'}{\displaystyle L}\right)'\right) \\
H_r&=&P_RR'-LP_L'.  \end{array}
\end{equation}
where $N, N^r, \hat N$ and $\hat N^r$ are Lagrange multipliers in the 
Hamiltonian formalism.
The system of constraints contain two surface constraints in
addition to usual Hamiltonian and momentum constraints of the ADM formalism.

ADM constraints:
\begin{equation}
\label{constraints1}
\left\{
\begin{array}{l}
H=0\\
\\
H_r=0\\
\end{array}
\right.
\end{equation}

Shell constraints:
\begin{equation}
\label{shellconstraints}
\left\{
\begin{array}{l}
\hat H_r=\hat\pi+\hat L\left[ P_L\right] =0\\
\\
\hat H=\frac{\displaystyle R\left[ R'\right] }{\displaystyle GL}+
\sqrt{M^2+\left.\hat\pi^2\right/ L^2}=0
\end{array}
\right.
\end{equation}

Karel Kuchar proposed some specific canonical transformation of 
the variables
$(R, P_R, L, P_L)$ to new canonical set $(R, \bar P_R, M, P_M)$ in
which Hamiltonian and momentum constraints given by (\ref{constraints})
are equivalent to the very simple set of constraints :
\begin{equation}
\label{simple}
\begin{array}{l}
\bar P_R=0\\
\\
M'=0
\end{array}
\end{equation}
The idea is to use the Schwarzschild anzatz for the space-time metric 
instead of the metric (\ref{metric}):
\begin{equation}
\label{schwar}
ds^2=-F(R,m) dT^2+\frac{\displaystyle 1}{\displaystyle F(R,m)} dR^2+
R^2 (d\theta^2+\sin^2\theta d\phi^2)
\end{equation}
where $T, R$ and $m$ are some functions of $(r,t)$ and $F(R,m)=1-\left. 2Gm
\right/ R$, and m, in general, is a function of $r$, $m=m(r)$.
 Equating the two forms of the metric (\ref{metric}) and
(\ref{schwar}) we obtain the transformation between the two sets of
dynamical variables. The explicit form of the transformation is
\begin{equation}
\label{transformation}
\begin{array}{rcl}
L&=&\sqrt{ \frac{\displaystyle (R')^2}{\displaystyle F}-FP_m^2}\\
P_L&=&\frac{\displaystyle RFP_m}{\displaystyle G}\sqrt{\frac{\displaystyle
(R')^2}{\displaystyle F}-FP_m^2}\\ R&=&R\\ \bar
P_R&=&P_R+\frac{\displaystyle P_m}{\displaystyle 2G}+ \frac{\displaystyle
FP_m}{\displaystyle 2G}+ \frac{\displaystyle
(RFP_m)'RR'-RFP_m(RR')'}{\displaystyle GRF\left( (R')^2\left.  \right/
F-FP_m^2\right) }
\end{array}
\end{equation}
where $P_m=-T'$. The Liouville form
\begin{equation}
\label{liuvil}
\Theta=\int P_R \dot R+P_L \dot L
\end{equation}
can be expressed in the new variables as follows:
\begin{eqnarray}
\label{important}
\Theta&=&\int P_m\dot m +\bar P_R\dot R+\frac{\displaystyle \partial}{
\displaystyle \partial t}\left( LP_L+
\frac{\displaystyle 1}{\displaystyle 2G}RR'\ln{\left|
\frac{\displaystyle RR'-LP_LG}{\displaystyle RR'+LP_LG}\right|}\right)
\nonumber \\
&+&
\frac{\displaystyle \partial}{\displaystyle \partial r}\left(
\frac{\displaystyle 1}{\displaystyle 2G}R\dot{ R}\ln{\left|
\frac{\displaystyle RR'+LP_LG}{\displaystyle RR'-LP_LG}\right| }\right).
 \end{eqnarray}
When there is no shell the  total derivatives in (\ref{important}) give 
rise to some surface terms at infinities.
 As was shown by K.Kuchar the appropriate falloff conditions
at infinities make the last  surface  term in the Eqn.(\ref{important}) zero. 
Then it follows from (\ref{important}) that  $(R, \bar P_R, m, P_m)$ form 
the canonical set of variables, and Eqn.(\ref{transformation}) 
describes a canonical transformation between   $(R, P_R, L, P_L)$ and 
 $(R, \bar P_R, m, P_m)$ .

In the presence of the thin shell the situation is different. Now surface 
the terms should not be neglected.
Let us do the transformation (\ref{transformation}) in regions I and II of
our space-time.The Liouville form of our Hamiltonian system (\ref{hamil})
has the form
\begin{equation}
\tilde \Theta =\int\limits_{I+II} P_R\dot R+P_L\dot L+\int\limits_\Sigma
\hat\pi\dot{\hat r}.
\end{equation}
After integration the total derivatives in (\ref{important}) give some 
contribution to the Liouville form on $\Sigma$:
\begin{eqnarray}
\tilde \Theta &=&\int\limits_{I+II} \bar P_R\dot R+P_m\dot m+
\int\limits_\Sigma
\left[LP_L+
\frac{\displaystyle 1}{\displaystyle 2G}
RR'\ln{\left|
\frac{\displaystyle RR'-LP_LG}{\displaystyle RR'+LP_LG}\right|} \right]
\dot{\hat r} dt
\nonumber \\
&-&\int\limits_\Sigma \left[
\frac{\displaystyle 1}{\displaystyle 2G}R\dot R\ln{\left|
\frac{\displaystyle RR'+LP_LG}{\displaystyle RR'-LP_LG}\right| }\right]+
\int\limits_\Sigma \hat\pi\dot{\hat r}
\nonumber \\
&=&
\int\limits_{I+II}P_m\dot m+\bar P_R\dot R+\int\limits_\Sigma \hat p\dot
{\hat r} + \int\limits_\Sigma \hat P_{\hat R} \dot{\hat R}
\end{eqnarray}
where we denoted
\begin{equation}
\label{newtransf}
\begin{array}{rcl}
\hat p&=&\hat\pi+L\left[ P_L\right]\\
\\
\hat P_{\hat R}&=& \pm \left[ \frac{\displaystyle 1}{\displaystyle 2G}
R\ln{\left|
\frac{\displaystyle RR'-GLP_L}{\displaystyle RR'+GLP_L}\right|} \right]
\end{array}
\end{equation}
and made use of the identity
\begin{equation}
\label{rhdot}
\dot{\hat R}=
\frac{\displaystyle d}{\displaystyle dt} R(t,\hat r(t))=
(\dot R(t,r)+R'(t,r)\dot{\hat r}(t))\left.\right|_{r=\hat r(t)}
\end{equation}
The sign in front of logarithm in the definition of the momentum $P_{\hat R}$ 
depends on whether we intersect the shell from ``in'' to ``out'' when 
going along the time coordinate curve from past to future (``+''-sign) or 
the other way around (``-''-sign). 
We see that this canonical transformation involves all the set of 
coordinates in the phase space 
$\Pi=\{  (R(r,t), P_R(r,t), L(r,t), P_L(r,t), \hat {r}(t),
\hat{\pi}(t)) \}$
 according to the formulae (\ref{transformation}) and (\ref{newtransf}).
Moreover it introduces additional pair of canonically conjugate variables
$(\hat R, \hat P_{\hat R})$ on the shell.
In both inner and outer regions I and II constraints are simplified due to 
the canonical transformation as it was in the absence of the shell 
(\ref{simple}). The surface momentum constraint $\hat H_r=0$ 
(\ref{constraints}) takes the form
\begin{equation}
\label{exclude}
\hat p=0
\end{equation}

To go further we need to know the Hamiltonian constraint on the shell in 
terms of these new canonical variables. To do this, let us consider more 
carefully Eqn.(\ref{rhdot}) which is the full time derivative of the shell 
radius. Using the definition $P_L=-\frac
R{GN}\left( \dot R-R^{\prime }N^r\right) $ we get:
\begin{equation}
\label{dotR}
\dot {\hat R}=-\frac{GNP_L}R+R^{\prime }\left( \dot r+N^r\right)
\end{equation}
Remembering that
$$
\pi \equiv \frac{ML^2\left( N^r+\dot r\right) }{\sqrt{N^2-L^2\left( N^r+\dot
r\right)^2}} 
$$
we can find
$$
L\left( N^r+\dot r\right) =\frac{\pi N}{L\sqrt{M^2+
\frac{\pi ^2}{L^2}}}
$$
Eqn.( \ref{dotR}) now reads
\begin{equation}
\frac{\dot {\hat R}}N=-\frac{G P_L}R+\frac{%
R^{\prime }}{L}\frac \pi {L\sqrt{M^2+\frac{\pi ^2}{L^2}}}
\end{equation}
It is now easy to see that the jump of $\dot{\hat R}$ across the shell is a 
linear combination of constraints
\begin{equation}
\left[ \dot {\hat R}\right] =\frac {GN}R \left( \chi H -\frac{H^r}L\right)
\end{equation}
where
$$
\chi =\frac{\pi}{L\sqrt{M^2+\frac{\pi ^2}{L^2}}}
$$
Now, the definition of the momentum $\dot{\hat R}$ can be  rewritten as follows
\begin{equation}
\label{beta}
\beta \equiv {\large e}^{\pm \frac{\displaystyle G\hat{P}_R}{\displaystyle R}}
\sqrt{\frac{F_{out}}{F_{in}}}=
\frac{\left( \frac{\dot{\hat R}}N+G\frac{P_L}R\left( 1+\chi
\right) \right) _{in}}{\left( \frac{\dot{\hat R}}N+G\frac{P_L}R\left( 1+\chi
\right) \right) _{out}}\equiv \frac{\alpha +y_{in}\left( 1+\chi \right) }{%
\alpha +y_{out}\left( 1+\chi \right) }
\end{equation}
where $\alpha =\frac{\dot{\hat R}}N$ and $y=G\frac{P_L}R$.\\ 
The next step is to find  the relation between $\alpha$ and $y$. 
From the definitions of $P_L$ and $\dot{\hat R}$ we have
$$
\frac {{R^{\prime }}^2}{L^2}=F+y^2.=\left(\frac \alpha \chi +\frac y\chi\right)
$$.
Solving this  for $y$ we get
\begin{equation}
\label{y}
y=\frac{\alpha \pm \sqrt{\alpha ^2-\left( \chi ^2-1\right) \left( F\chi
^2-\alpha ^2\right) }}{\chi ^2-1}
\end{equation}
Substituting this expression back into Eqn. (\ref{beta}) we obtain 
after simple algebraic transformations 
\begin{eqnarray}
\label{hamilton}
&\beta& =\frac{z-\sigma _{in}\sqrt{z^2+F_{in}}}{z-\sigma _{out}
\sqrt{z^2+F_{out}}}\\
\mbox{where} &z^2&=\frac{\alpha ^2}{1-\chi ^2}
\end{eqnarray}
Making use of the momentum constraint we can wright the jump of $y$ as 
follows 
\begin{equation}
\begin{array}{c}
\sigma _{out}\chi
\sqrt{\alpha ^2+\left( 1-\chi ^2\right) F_{out}}=\sigma _{in}\chi \sqrt{%
\alpha ^2+\left( 1-\chi ^2\right) F_{in}}-G\frac MR\chi \sqrt{1-\chi ^2}%
  \\  \\
\end{array}
\end{equation}
Note that this is just our starting equation (\ref{emotion}) if we choose 
the proper time (substituting $\dot {\hat R}^2$ for $z^2$.)
Squaring this equation  we get
\begin{equation}
\left\{
\begin{array}{c}
\sigma _{in}
\sqrt{z^2+F_{in}}=-\frac{R\left[ F\right] }{2MG}+\frac {MG}{2R} \\  \\
\sigma _{out}\sqrt{z^2+F_{out}}=-\frac{R\left[ F\right] }{2MG}-
\frac {MG}{2R}
\end{array}
\right.
\end{equation}
\begin{equation}
z=\pm\sqrt{\left( \frac{R\left[ F\right] }{2MG}\right) ^2-\frac 12\left(
F_{out}+F_{in}\right) +\frac{M^2G^2}{4R^2}}
\end{equation}
The sign for $z$ is opposite to the sign for $\dot{\hat R}$. Finally we get 
for the shell constraint
\begin{equation}
\label{frac}
\beta - \frac{z+\frac{R\left[ F\right] }{2MG}-\frac {MG}{2R}}{z+\frac{R\left[
F\right] }{2MG}+\frac {MG}{2R}}=0
\end{equation}

It can be easily shown that the above constraint is equivalent to 
the Hamiltonian constraint (\ref{hamilton}).
In what follows we will consider the following squared version of the 
Hamiltonian constraint (\ref{hamilton}) as the suitable classical counterpart
for the quantum constraint for the wave function $\Psi$
\begin{equation} 
\label{Cons}
C=F_{out}+F_{in}-\sqrt{F_{out}}\sqrt{F_{in}}\left[ \exp \left(
\frac {G\dot P_R} {R} \right)+
 \exp \left(- \frac {G\dot P_R} {R}\right)\right] - \frac{M^2G^2}{R^2}
\end{equation}

 The Hamiltonian constraint (\ref{hamilton}) was derived under the assumption 
that both $F_{in}$ and $F_{out}$
are positive . It is possible, of course to derive analogous constraints in $T_\pm$-regions, 
where $F<0$. But, instead, we make the following substitution
\begin{equation}
\sqrt{F}\rightarrow F^{1/2}
\end{equation}
and consider this function as a function of complex variable. Then the point of the horizon
$F=0$ becomes a branching point , and we need the rules of the bypass. We assume the following 
\begin{equation}
\label{bypass2}
\begin{array}{rl}
F^{1/2}&=\left| F\right| e^{i\phi}\\
\phi =0& \mbox{\ in\ } R_+\mbox{-region}\\
\phi = \pi /2& \mbox{\ in\ } T_-\mbox{-region}\\
\phi =\pi & \mbox{\ in\ } R_-\mbox{-region}\\
\phi =-\pi /2& \mbox{\ in\ } T_+\mbox{-region}\\  
\end{array}
\end{equation}
for the black hole case , and 
\begin{equation}
\label{bypass3}
\begin{array}{rl}
\phi =\pi& \mbox{\ in\ } R_+\mbox{-region}\\
\phi = -\pi /2& \mbox{\ in\ } T_-\mbox{-region}\\
\phi =0 & \mbox{\ in\ } R_-\mbox{-region}\\
\phi =\pi /2& \mbox{\ in\ } T_+\mbox{-region}\\  
\end{array}
\end{equation}
for the wormhole case.
The reason for  such analytical continuation is that we are able to get the single
equation on the wave function $\Psi$ which covers all four patches of the complete
Penrose diagram for the Schwarzschild spacetime.
The Carter-Penrose diagram for $\sigma=\pm1$ (balck hole and worm hole case,
correspondingly) are shown in Fig.10.

\begin{figure}[htbp] 
\vspace*{13pt}
\centerline{\includegraphics{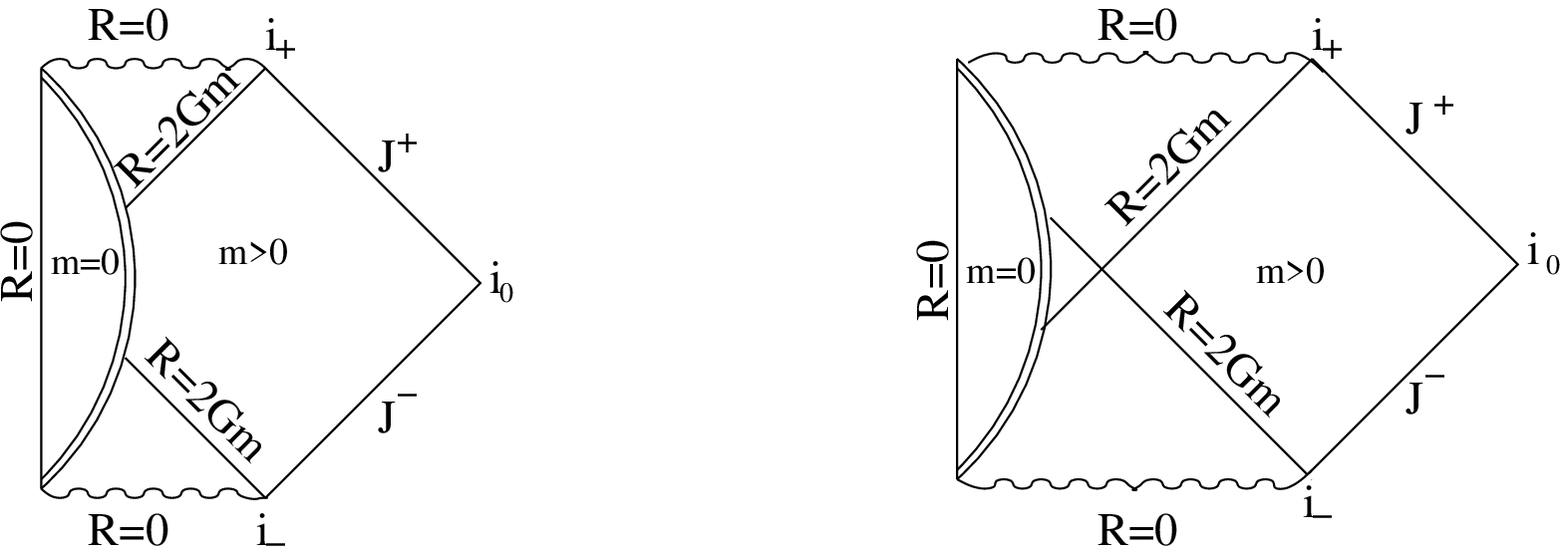}}
\vspace*{13pt}
\caption{}
\end{figure}

It should be stress here that we consider the thin selfgravitating shell together
with the space-time which it ``creates''. Thus for different type of shell's 
motion we have to choose different forms of Hamiltonian constraints. For the 
shell escaping to infinity we choose the following constraint (see Fig.8 for
the corresponding Carter-Penrose diagram in the case $F_{in}=1$)

\begin{equation}
\label{outs}
e^{\pm \frac{G p}{R}} = \frac{\frac{1}{2} (F_{out} + F_{in}) - 
\frac{M^2 G^2}{2 R^2} \pm \frac{M G}{R} z}{\sqrt{F_{in}} \sqrt{F_{out}}} , 
\end{equation}

while for the falling shell that start from infinity (Carter-Penrose diagram
is shown in Fig.9) we have

\begin{equation}
\label{ins}
e^{\pm \frac{G p}{R}} = \frac{\frac{1}{2} (F_{out} + F_{in}) - 
\frac{M^2 G^2}{2 R^2} \mp \frac{M G}{R} z}{\sqrt{F_{in}} \sqrt{F_{out}}} , 
\end{equation}

The bound motion is symmetric to the change of the momentum sign,so for the
Hamiltonian constraint we choose the sum of the constraints (\ref{outs}) and
(\ref{ins}), namely

\begin{equation}
\label{inout}
e^{\frac{G p}{R}} + e^{-\frac{G p}{R}} = 
\sigma \frac{F_{out} + F_{in} - \frac{M^2 G^2}{R^2}}{\sqrt{F_{in}} 
\sqrt{F_{out}}} 
\end{equation}

We now turn to the Dirac constraint quantization procedure . 

It is convenient to make a canonical transformation from
 $(\hat R, \hat P_{\hat R})$
to $(\hat S, \hat P_S)$:
\begin{equation}
\label{area}
\left\{
\begin{array}{rcl}
\hat S&=&=\frac{\displaystyle \hat R^2}{\displaystyle R_0^2}\\
\hat P_S&=&R_0^2\frac{\displaystyle \hat P_{\hat R}}{\displaystyle 2R}
\end{array}
\right.
\end{equation}
where $R_0$ is some value radius of the shell. 
Dimensionless variable $\hat S$ is the surface area of 
the shell measured in the units of horizon  area of the shell of mass $M$.  

The phase space of our model consist of coordinates 
$(R(r), \tilde P_R(r), m(r),\\
 P_m(r), \hat S, \hat P_S, \hat r, \hat p_r)$
$r\in (-\infty, \hat r-\epsilon )\bigcup (\hat r+\epsilon, \infty)$.
Then the wave function in coordinate representation depends on configuration space coordinates:
\begin{equation}
\label{wave}
\Psi=\Psi(R(r), m(r), \hat S, \hat r)
\end{equation}
and all the momenta become operators of the form
\begin{equation}
\label{operators}
\begin{array}{cc}
\tilde P_R(r)=-i\left.\delta\right/\delta R(r)& P_m(r)=-i\left.\delta\right/
\delta m(r)\\
\hat P_S=-i\left.\partial\right/\partial \hat S& \hat p_r=-i\left.\partial\right/\partial \hat r
\end{array}
\end{equation}
ADM and shell constraints (\ref{constraints1}) and (\ref{shellconstraints}) become operator equations on
$\Psi$. The set of ADM constraints is equivalent to the set of constraints (\ref{simple}) in Kuchar variables
which could be easily solved on quantum level. Indeed, in the regions I and II the equations
\begin{equation}
\label{qkuchar}
\left\{
\begin{array}{rcl}
\left.\partial \Psi\right/\partial R(r)&=&0\\
M'(r)\Psi&=&0
\end{array}
\right.
\end{equation}
express the fact that wave function does not depend on $R(r)$ and the dependence on $M(r)$ is
reduced in each region I and II to $\Psi\equiv \delta(M-M_{\pm})$ where $M_{\pm}$ defined
in the regions I (-) and II (+) do not depend on $r$. They equal to Schwarzschild masses 
in the inner and outer regions $M_{in}$ and $M_{out}$ in (\ref{main}).

The set of shell constraints (\ref{shellconstraints}) impose further restrictions on $\Psi$. 
First of them takes the form
\begin{equation}
\left.\partial \Psi\right/\partial \hat r=0
\end{equation}
 in new variables according to (\ref{exclude}). So the only nontrivial equation is the shell
constraint (\ref{Cons}) (or (\ref{hamilton},(\ref{frac}),(\ref{main})which
are classically equivalent to the Eqn.(\ref{Cons}))
\begin{eqnarray}
\hat  C \left(m_+,m_-,\hat S, -i\hbar \partial /\partial \hat S\right)=0 \\
\nonumber
\Psi = \Psi (m_+,m_-,\hat S)
\end{eqnarray}
The operator $\hat C$ contains the exponent of the of the momentum $\hat P_S$. 
This exponent becomes an operator  of finite displacement when $\hat P_S$
becomes differential operator:
\begin{equation}
\label{finite}
e^{\frac{G \hat P_R}{R}}=e^{\frac { \hat P_S}{R_0^2}}\Psi=
e^{-i\zeta \frac{\partial}{\partial \hat S}}\Psi =
\Psi(m_+,m_-,\hat S -\zeta i)
\end{equation}
$\zeta =\frac {2G}{R_0^2}$

For the sake of simplicity we consider here only the case of bound motion of the 
single shell having the bare mass $M$. It means that $m_{in}=0$ and we should be 
use the symmetric form Hamiltonian constraint. In this case it is naturally to 
choose for $R_{0}$ the Schwarzschild radius $R_0=R_g=2Gm$ ($m=m_{out}$). For 
$\zeta$ we have now $\zeta =\frac {1}{2} (\frac {m_{pl}}{m})^2 $,and the Hamiltonian
constraint can be written in the form

\begin{equation}
\label{fors}
\Psi ({S + i \zeta) + \Psi(S - i \zeta}) = 
\frac{2 - \frac{1}{\sqrt{S}} - \frac{M^2}{4 m^2 s}}
{(1 - \frac{1}{\sqrt{S}})^{\frac{1}{2}}} \Psi(S) . 
\end{equation}

We have mentioned already  that the classically equivalent constraints  give
inequivalent quantum theories. This is well known fact. We suggest that the 
criterion to choose the correct  quantum theory is the  behavior of the wave 
functions in the quasiclassical regime.In our case this means the large black 
holes limit. Indeed , the parameter $\zeta =\frac {1}{2} (\frac {m_{pl}}{ m})^2 $
becomes small for large masses , and the expansion with respect to this 
parameter is equivalent to the expansion in Planckian constant $\hbar$ 
($m_{pl}=\sqrt{\hbar / Gc}$). In the next Section we will consider this
quasiclassical limit and show that our choice for the quantum constraint 
is a good one (at least in the case of one thin shell).
At the and of this Section we would like to make an important remark.
Our quantum equation \ref{fors} (which is just a Schroedinger equation )
is the equation in finite differences rather than differential equation,
and the shift in argument is along an imaginary axis. In the case of 
differential equation we we require the solution to be differentiable 
sufficiently many times . Similarly, we have to demand the solutions of
our finite differences equation \ref{fors} to be analytical functions.
This condition is very restrictive but unavoidable. Our previous experience
(see \cite{20}) shows that it is the analyticity of the wave functions 
and not the boundary conditions that lead to the existence of the discrete
mass (energy) spectrum for bound states.

 The finite difference equation (\ref{fors}) becomes an ordinary differential equation in quasiclassical
limit which is the same as the limit of large ( $m\gg m_{pl}$ ) black holes. Indeed the parameter of
finite displacement of the argument of $\Psi$ in (\ref{fors})
  $\zeta$
becomes small and we could cut the Tailor  expansion
\begin{equation}
\Psi (\hat S+\zeta i)=\Psi (\hat S)+i\zeta \Psi'(\hat S)-
\frac{\displaystyle \zeta^2}{\displaystyle 2}\Psi''(\hat S)+...  
\end{equation}
at the second order. Thus,

\begin{equation}
\label{expansion}
\begin{array}{rcl}
\left. \Psi\right|_{S\pm \zeta i}&\approx&\Psi(S)\pm \Psi'(S)\zeta i-
\frac{\displaystyle \zeta^2}{\displaystyle 2}\Psi''(S)\dots
\\
\\

\left. F^{\frac{1}2}\right|_{S\pm\zeta i}
&=&\sqrt{1-\frac{\displaystyle 1}{\displaystyle \sqrt{s\pm \zeta i}}}\approx 
F^{\frac{1}2}
\left( 
1\pm
\frac{\displaystyle 1}{\displaystyle 2FS^{3/2}}\zeta i+
\left(
\frac{\displaystyle 3}{\displaystyle 8FS^{5/2}}+
\frac{\displaystyle 1}{\displaystyle 8F^2S^3}
\right)
\zeta^2
\right) \dots\\ 
\\
 
\end{array}
\end{equation} 
This leads to ordinary differential equations of second order, which are 
different in $ R_+, R_-, T_+ $ and $T_-$ regions
due to the different values of the phases  in Eqn.(\ref{fors}). 
The interesting for us singular points of these differential equations are
\begin{equation}
S=\infty  \mbox{\ and\ } S=1.
\end{equation}
In the quasiclassical limit our requirement of the analyticity of the solutions
 to  the exact equation (\ref{fors}) transforms into the requirement that the
branching points of the leading terms in the solutions to the approximate
equations should be of the same kind . Thus, we need to keep only those terms
in the corresponding equations that give us these leading terms. Below we 
consider  the black hole case only. The results are easily translated to the 
wormhole case. 

The singular point $S=\infty$ in the region $R_+$ lies in a classically 
forbidden region as far as we restrict ourselves with bound motions of 
the shell only. In order to analyze the behavior of $\Psi$ in this region
we should take (\ref{fors}) with $\phi =0$ and expand all the quantities 
in terms of $y$, where $s=(1+y)^2$
The result is   
\begin{equation}
\label{infty}
\Psi_{yy}-\frac{1}y \Psi_{y}+ \frac{1}{\displaystyle \zeta^2}
\left( 1-\frac{\displaystyle M^2}{\displaystyle m^2}+
\frac{1}{2y}(2-\frac{\displaystyle M^2}{\displaystyle m^2})\right)\Psi=0
\end{equation}
The leading term of the solution is
\begin{eqnarray}
\Psi &\sim &
y^{\displaystyle \frac{1}{2}-\frac
{\displaystyle \frac{\displaystyle M^2}{\displaystyle m^2}-2}
{\displaystyle 4\mu\zeta^2}}
\exp(-\mu y), \\
\nonumber
\mu&=&\frac{1}{\zeta} 
\sqrt{\frac{\displaystyle M^2}{\displaystyle m^2}-1},
\ y\gg\zeta
\end{eqnarray}

For another singular point in $R_+$ region , that is for $S\rightarrow 1+0$
we have $(s=(1+z^2)^2)$
\begin{equation}
\label{z}
\Psi_{zz}- 3z\Psi_{z}+ \frac{16z}{\displaystyle \zeta^2}
\left( 1-\frac{\displaystyle M^2}{\displaystyle 4m^2}\right)\Psi=0
\end{equation}
with leading term 
\begin {eqnarray}
\Psi \sim 
1-\frac{8}{\displaystyle 3\zeta^2}
\left( 1-\frac{\displaystyle M^2}{\displaystyle 4m^2}\right)y^{3/2}\\
\nonumber
y=\sqrt{z}, \ s\gg\zeta,\ y\gg\zeta,\ \ \zeta\ll 1
\end{eqnarray}

Comparing the types of the branching points at $s\rightarrow\infty$
and $s\rightarrow1+0$ we can conclude that 
\begin{equation}
\label{n}
\frac
{\displaystyle 2-\frac{\displaystyle M^2}{\displaystyle m^2}}
{\displaystyle 4\zeta\sqrt{\frac{\displaystyle M^2}{\displaystyle m^2}-1}}
=n, \ \ n=\mbox{integer}
\end{equation}
This is the first quantization condition. We will not consider here the 
wormhole case . Note only that , as can be shown , the positive values 
of quantum number n correspond to black holes while negative n correspond 
to wormholes.
 
We do not consider here separately the asymptotics  in $R_-$ region near the
horizon $(s\rightarrow1+0)$ because it differs from the corresponding solution 
in $R_+ $-region only by the sign in front of the second term.

Let us now turn to the asymptotics of the solutions in $R_-$-region for 
$s\rightarrow\infty$. Due to the minus sign in front of $F^{1/2}$ the 
equation for the wave function in a $R_-$-region is quite different from
that in a $R_+$-region 

\begin{equation}
\label{yy}
\Psi_{yy}-\frac{1}y \Psi_{y}- \frac{1}{\displaystyle \zeta^2}
\left( 16y^2+1-\frac{\displaystyle M^2}{\displaystyle m^2})\right)\Psi=0
\end{equation}
The leading term of the asymptotic is now the following 
\begin{equation}
\Psi \sim 
y^{\displaystyle \frac
{\displaystyle \frac{\displaystyle M^2}{\displaystyle m^2}-1}
{\displaystyle 8 \zeta}}
\exp -(\frac {2}\zeta y^2)
 \end {equation}
Note that the falloff in the $R_-$-region is much faster than it is in the 
$R_+$-region. This is a quite reasonable  result because it means that the
 quantum shell in the black hole case can penetrate into the $R_-$-region 
(which is completely forbidden for the classical motion) but the probability
 of such  an event is negligible small.

And , again, comparing the types of the branching points at $s\rightarrow1+0$
and $s\rightarrow\infty$ in the $R_-$-region we get 
   
\begin{equation}
\label{p}
{\displaystyle \frac
{\displaystyle \frac{\displaystyle M^2}{\displaystyle m^2}-1}
{\displaystyle 8 \zeta}}=\frac{1}2+p, \qquad p=\mbox{positive}
 \quad \mbox{integer}
 \end {equation}
The appearance of the second quantum number is rather surprising and its
origin is in nontrivial casual structure of the Schwarschild space-time.
The classical shell in the (e.g.) black hole case cannot move in the 
$R_-$-region, and simply absent on the complete Carter-Penrose diagram.
But the quantum shell with parameters corresponding to the black hole case,
``feel'' both $R_+$- and $R_-$-regions, this why the second quantum number
plays its role. It is instructive to compare our result with the famous
Coulomb problem for the hydrogen atom. The parameters of the Coulomb problem
are the proton's and lectron's electric charge and all of them being
quantized from very begining. Therefore, to obtain a discrete spectrum
for bound states it is enough to have only one quantum condition. But, in
our case there are two continous parameters, the total mass $m$ and 
bare mass $M$. And now, in order to obtain a discrete mass (energy) spectrum,
we should have two quantum conditions. And just what we really got!

Now, we would like to consider the behavior of the
solutions in the vicinity of the horizon (sub-Planckian deviation ), where
$|y|\gg \zeta $ $(s\sim 1)$. To be specific we will be interested in the 
solutions $R_+$ and $T_-$ regions. The expansion (\ref{expansion}) is no  more 
 valid for
the function  $F^{1/2}(s\pm i\zeta)$ but it is still valid for the wave $\Psi$.
Keeping the leading terms only we have now 
\begin{equation}
\label{s}
\Psi_{ss}(s)- \frac{2}{\displaystyle \zeta}\Psi_{s}(s)
+\left( \frac{4}{\displaystyle \alpha\zeta^{5/2}}
 (1-\frac{\displaystyle M^2}{\displaystyle 4m^2})
-\frac{2}{\displaystyle \zeta^2}
\right)\Psi(s)=0
\end{equation}
with the solution 
\begin{equation}
\Psi\sim e^{ks}, \quad k\approx - \frac{1}\zeta \pm
\sqrt{- \frac{4}{\displaystyle \alpha\zeta^{5/2}}
 (1-\frac{\displaystyle M^2}{\displaystyle 4m^2})} 
\end{equation}
The coefficient $\alpha$ equals to 1 in the $R_+$-region and to imaginary
unit i in the $T_-$-region.

In the $R_+$ region 
\begin{equation}
k\approx - \frac{1}\zeta \pm
i\sqrt{- \frac{4}{\displaystyle \zeta^{5/2}}
 (1-\frac{\displaystyle M^2}{\displaystyle 4m^2})} 
\end{equation}
and we have superposition of two waves (ingoing and out outgoing) with
relatively equal amplitudes.

in the $T_-$-region
\begin{eqnarray}
k\approx - \frac{1}\zeta \pm
\sqrt{- \frac{4i}{\displaystyle \zeta^{5/2}}
 (1-\frac{\displaystyle M^2}{\displaystyle 4m^2})} =
 - \frac{1}\zeta \pm
 \frac{\sqrt{2}}{\displaystyle \zeta^{5/4}}
 \sqrt{(1-\frac{\displaystyle M^2}{\displaystyle 4m^2})}(1+i)= \\
\nonumber
\left(\pm  \frac{\sqrt{2}}{\displaystyle \zeta^{5/4}}
 \sqrt{(1-\frac{\displaystyle M^2}{\displaystyle 4m^2})}-\frac{1}\zeta)\right)
\pm i  \frac{\sqrt{2}}{\displaystyle \zeta^{5/4}}
 \sqrt{(1-\frac{\displaystyle M^2}{\displaystyle 4m^2})}
\end{eqnarray}
The existence of two waves in the $T_-$-region reflects the quantum trembling
of the horizon . But the outgoing wave is enormously damped relative to the 
ingoing wave (of course , in the $T_-$- region the situation is exactly 
inverse one). It is this damping that cause (in a quasiclassical regime)
existence of the single ingoing wave in the $T_-$-region at the distances
larger than Planckian.

In the end of this Subsection let us discuss briefly some features of the 
obtained discrete mass spectrum (\ref{n}) and (\ref{p})

Let us discuss some properties of the spectrum that arises from these 
conditions. 

1. For larger values of quantum number $n$ ($\frac{M^2}{m^2}-1<<0$) one 
can easily derive nonrelativistic Rydberg formula for Kepler's problem, 
$E_{nonrel} = M-m = -\frac{G^2 M^4}{8 n^2}$. 

2. The role of turning point $\rho_0$ is now played by the integer $n$. 
Thus, keeping $n$ constant and calculating 
$\gamma =\frac{\partial m}{\partial M} | _n $ 
one can distinguish 
between a black hole case ($\gamma > 0$) and a wormhole case ($\gamma < 0$). 
It appears that 
$ \frac{\partial m}{\partial M} \vert _n  >0$ for  $n \ge n_0$, 
negative or zero, and 
\begin{equation}
\vert n_0 \vert = E [\sqrt{2} \sqrt{13\sqrt{5}-29} (1 + 2p)]
\label{6}
\end{equation}
where $E[\cdots]$ means the ``integer part of ...''. It is interesting, that
for $n \ge 0$
 the asyptotical behaviour  of the wave function is exactly the 
same as in the Coulomb problem, but for the negative values of $n_0$ there is
no such analogy.
3. There exists a minimal possible value for a black hole mass. 
This occurs if $p = n_0 = 0$,
\begin{equation}
m_{min} = \sqrt {2} m_{pl}
\label{7}
\end{equation} 

4.It is interesting to study the limiting case when  
$M \to \infty$ and total mass $m$ is kept finite. It is correspond to the 
limit of eternal black hole. From our spectrum it easy to deduce that 
\begin{equation}
\label{metern}
m = \lim_{n,p \rightarrow \infty} \left(\frac{n}{\sqrt{p}} \right) m_{Pl} ,
\end{equation}  
and, as result, we obtain that mass of the black hole may be arbitrary
(but positive).

5. The our spectrum is not universal in the sense that corresponding 
wave function form two-parameter family $\Psi_{n,p}$.
But for quantum Schwarzschild black hole we expect a one-parameter 
family of wave functions. Quantum black holes should have no 
hairs, otherwise there will be no smooth limit to the classical 
black holes. All this means that our spectrum is not a quantum 
black hole spectrum, and our shell does not collapse (like an 
electron in hydrogen atom). Physically, it is quite understandable, 
because the radiation is yet included into consideration. 
But before doing this we would like to make one important note.

Till now we considered only the case of zero mass $m_{in}$ inside 
the shell. The generalization to the general case is not difficult,
and result is the following

\begin{eqnarray}
\label{specex}
\frac{2 (\Delta m)^2 - M^2}{\sqrt{M^2 - (\Delta m)^2}} = 
\sqrt{2 m_{Pl}^2} n ,  \nonumber \\
\\ \nonumber
M^2 - (\Delta m)^2 = 2 m_{Pl}^2 (1 + 2 p) ,
\end{eqnarray}

Where $M$ is the bare mass of the shell, $m = m_{out}$ is the total mass
of the system, and $\Delta m = m_{0ut} - m_{in}$. Moreover this spectrum
is,in fact, the exact spectrum of our finite difference equation.
The reason why the formulas obtained for large black holes appeared the 
exact ones is pure mathematical and we will not discuss it here. But one 
important feature of the exact spectrum (\ref{specex}) should be 
mentioned. We know already that all linear homogeneous equation in finite 
differences obey the following feature, namely, any its solutions 
multiplied by the periodic function which period  equals to the argument
shift of the equation, gives again the solution. This leads, in general,
to infinite degeneracy of the spectrum. But our spectrum (\ref{specex})
can be made nondegenerate. The matter is that as the argument of our 
equation in finite differences (\ref{fors}) we have square of the radius and
leading term of the asyptoticalsolution at the right infinity 
(in $R_+$-region), in general, has a form $\Psi \sim e^{\alpha r^2}$, but
there is one exception. The exceptional solution has an asymptotic
$\Psi \sim e^{\lambda r}$, and just the same asymptoticly behavior exhibit
the wave function of the nonrelativistic Coulomb problem. Thus if we demand
the existence of the ``correct'' non-relativistic limit, then ``correct''
wave function should have the leading asymptotical term
$\Psi \sim e^{\lambda r}$, and such a requirement removes a degeneracy.

\subsection{Radiation and quantum collapse dynamics}

In order to investigate the quantum collapse, we should switch on the process
of quantum radiation. For the modelling of such radiation we again make use
 of the developed above theory of thin shells, 
but now shells are null and move with the
speed of light. The corresponding classical equation is our nonsymmetric
constraint (\ref{outs}) in which the bare mass $M$ is put zero. The 
quantum equation looks as follows \cite{30}
\begin{equation}
\Psi (m_{in}, m_{out}, s-i\zeta)=
\sqrt{\frac{1-\frac{\mu}{\sqrt{s}}}{1- \frac{1}{\sqrt{s}}}}
\Psi(m_{in}, m_{out},s)
\label{8}
\end{equation}
here $\mu = m_{in} / m_{out}$, $\zeta = \frac{1}{2} m_{pl} ^2 /m_{out} ^2$. 
The existence of the second infinity on the other side of the 
Einstein-Rosen bridge leads to the following quantization condition 
($m = m_{out}$)
\begin{equation}
\delta m = m_{out} - m_{in} = -2m+2 \sqrt{m^2+k m_{pl} ^2},
\label{9}
\end{equation}
where $k$ is an integer. It is interesting to note that if we 
put $k=1$ (minimal radiating energy) and require $\delta m <m$ 
(not more than the total mass can be radiated away), then we 
obtain 
\begin{equation}
m = m_{out} > \frac{2}{\sqrt{5}} m_{pl}.
\label{10}
\end{equation} 
Thus, the black hole with the mass given by  Eqn.(\ref{7}) is not 
radiating and, therefore, it can not be transformed into semi-closed 
world (wormhole-like case). 

The discrete spectrum of radiation (\ref{9}) is universal in the 
sense that it does not depend on the structure and mass spectrum 
of the gravitating source. This means that the energies of 
radiating quanta do not coincide with level spacing of the source. 
The most natural way in resolving such a paradox is to suppose 
that quanta are created in pairs. One of them is radiated away, 
while another one goes inside. Thus, the quantum collapse can not 
proceed without radiating even in the case of spherical symmetry. 
This radiation is accompanying with creation of new shells inside 
the primary shell we started with. We see, that the internal structure 
of quantum black hole is formed during the very process of quantum 
collapse. And if at the beginning we had one shell and knew everything 
about it, then already after the first pulse of radiation we have more 
than one way of creating the inner quantum. So, initially the entropy 
of the system was zero, it starts to grow during the quantum collapse. 
If somehow such a process would stop we would call the resulting 
object ``a quantum black hole''. The natural limit is the transition 
from black hole to the wormhole-like shell. The matter is that such 
a transition requires (at least in quasi-classical regime) insertion 
of an infinitely large volume, and the quasi-classical probability 
for this process is zero. 

Let us write down the spectrum of the shell with nonzero Schwarzschild 
mass, the total mass inside, $m_{in} \ne 0$
\begin{eqnarray}
\frac{2 (\Delta m)^2 - M^2}{\sqrt{M^2 - (\Delta m)^2}} = 
\frac{2 m_{pl} ^2}{\Delta m + m_{in}} n
\\
M^2 - (\Delta m)^2 = 2 (1 +2p) m_{pl} ^2 \nonumber
\label{11}
\end{eqnarray}
Here $\Delta m$ is the total mass of the shell, $M$ is the bare mass, 
the total mass of the system equals $m = m_{out}=\Delta m + m_{in}$. 
For the black hole case $M^2 < 4 m \Delta m$, or 
\begin{equation}
\frac{\Delta m}{M} > 
\frac{1}{2}(\sqrt{(\frac{m_{in}}{M})^2 +1} - \frac{m_{in}}{M}).
\label{12}
\end{equation}
After switching on the process of radiation governed by Eqn.(\ref{9}), 
the quantum collapse starts. Our computer simulations shows that evolves 
in the ``correct'' direction, e.g. it becomes nearer and nearer to the 
threshold (\ref{12}) between the black hole case and wormhole case. 
The process stops exactly at $n=0$!

The point $n=0$ in the spectrum is very special. Only in such a state 
the shell does not ``feel'' not only the outer regions (what is natural 
for the spherically symmetric configuration) but it does not know anything 
about what is going on inside. It ``feel'' only itself. Such a situation 
reminds the classical (non-spherical) collapse. Finally when all the 
shells (both the primary one and newly produced) are in the 
corresponding states $n_i = 0$, the system does not ``remember'' its 
own history. And this is a quantum black hole. The masses of all the 
shells obey the relation 
\begin{equation}
\Delta m_i = \frac{1}{\sqrt{2}} M_i.
\label{13}
\end{equation}
The subsequent quantum Hawking's evaporation can produced only via 
some collective excitations and formation, e.g., of a long chain of 
microscopic semi-closed worlds.

\subsection{Classical analog of quantum black hole}

Let us consider large ($m \gg m_{pl}$) quantum black holes. The 
number of shells (both primary ones and created during collapse) 
is also very large, and one may hope to construct some classical 
continuous matter distribution that would mimic the properties of 
quantum black holes. First of all, we should translate the 
``no memory'' state ($n=0$ for all the shells) into ``classical 
language''. To do this let us rewrite the Eqn.(\ref{emotion}) (energy 
constraint equation) for the shell, inside which there is 
some gravitating mass $m_{in}$, 
\begin{equation}
\sqrt{\dot {\rho }^{2}+1-\frac{2Gm_{in}}{\rho}}- \sqrt{\dot {%
\rho }^{2}+1-\frac{2Gm_{out}}{\rho }=}\frac{GM}{\rho }
\label{14}
\end{equation}
and consider a turning point, ($\dot \rho = 0$, $\rho = \rho _0$):
\begin{equation}
\Delta m = m_{out} - m_{in} = 
M \sqrt{1-\frac{2Gm_{in}}{\rho _0}} - \frac{G M^2}{2 \rho _0} .
\label{15}
\end{equation}
It is clear now that in order to make parameters of the shell 
( $\Delta m$ and $M$) not depending on what is going on inside 
we have to put $m_{in} = a \rho _0$.

Our quantum black hole is in a stationary state. Therefore, a 
classical matter distribution should be static. We will consider 
a static perfect fluid with energy density $\varepsilon$ and pressure 
$p$. A static spherically symmetric metric can be written as 
\begin{equation}
d s^2 = e^{\nu}dt^2-e^{\lambda}dr^2-r^2(d\theta^2+\sin^2\theta d \varphi ^2)
\label{16}
\end{equation}
where $\nu$ and $\lambda$ are functions of the radial coordinate 
$r$ only. The relevant Einstein's equations are (prime denotes 
differentiation in $r$)
\begin{eqnarray}
8\pi G\varepsilon = 
- e^{\lambda} (\frac{1}{r^2} - \frac{ \lambda '}{r})+ \frac{1}{r^2},
\nonumber
\\
-8\pi G p = 
- e^{\lambda} (\frac{1}{r^2} - \frac{ \nu '}{r})+ \frac{1}{r^2},
\\
-8\pi G p = 
- \frac{1}{2} e^{\lambda} 
(\nu'' +\frac{\nu'^2}{2} +\frac{\nu'-\lambda'}{r}-\frac{\nu'\lambda'}{2})
 \nonumber
\label{17}
\end{eqnarray}
The first of these equations can be integrated to yield 
\begin{equation}
e^{-\lambda} = 1- \frac{2Gm(r)}{r},
\label{18}
\end{equation}
where
\begin{equation}
m(r)=4\pi \int_0^r \varepsilon r'^2dr'
\label{19}
\end{equation}
is the mass function, that must be identified with $m_{in}$. 
Thus, $m(r)=ar$, and 
\begin{equation}
\varepsilon = \frac{a}{4\pi r^2},\qquad  e^{-\lambda} = 1-2Ga .
\label{20}
\end{equation}
We can also introduce a bare mass function 
\begin{equation}
M(r)=4\pi \int_0^r \varepsilon e^{\frac{\lambda}{2}}r'^2dr' ,
\label{21}
\end{equation}
and from Eqn.(\ref{20}) we get 
\begin{equation}
M(r)=\frac{ar}{\sqrt{1-2Ga}}
\label{22}
\end{equation}
The remaining two equations can now be solved for $p(r)$ and $e^{\nu}$. 
The solution for $p(r)$ that has the correct non-relativistic limit is 
\begin{equation}
p(r)=\frac{b}{4\pi r^2} ,\qquad 
b=\frac{1}{G}(1-3Ga-\sqrt{1-2Ga}\sqrt{1-4Ga}) ,
\label{23}
\end{equation}
and for $e^{\nu}$ we have 
\begin{equation}
e^{\nu} = C r^{2G \frac{a+b}{1-2Ga}} .
\label{24}
\end{equation}
The constant of integration $C$ can be found from matching of the 
interior and exterior metrics at some boundary $r=r_0$. Let us 
suppose that $r>r_0$ the space-time is empty, so the interior 
should be matched to the Schwarzschild metric. Of course, to 
compensate the jump in pressure ($\Delta p = p(r_0) = p_0$) we 
must introduce some surface tension $\Sigma$. From matching 
conditions (see, e.g. \cite{11}) it follow that 
\begin{eqnarray}
C = (1-2Ga) r_0 ^{-2G \frac{a+b}{1-2Ga}} ,
\nonumber
\\
e^{\nu} = (1-2Ga)( \frac{r}{r_0}) ^{2G \frac{a+b}{1-2Ga}} ,
\\
\Sigma = \frac{2 \Delta p}{r_0} = \frac{b}{2 \pi r_0 ^3}
 \nonumber
\label{25}
\end{eqnarray}

We would like to stress that the pressure $p$ in our classical 
model is not real  but only effective because it was introduce 
in order to mimic the quantum stationary states. We see, that the 
coefficient $b$ in Eqn.(\ref{23}) becomes a complex number if 
$a>1/4G$. Hence, we must require $a\le 1/4G$, and in the limiting 
point we have the stiffest possible equation of state $\varepsilon=p$ 
It means also that hypothetical quantum collective excitations 
(phonons) would propagate with the speed of light and could be 
considered as massless quasiparticles. It is remarkable that in the 
limiting point we have $m(r)=M(r)/\sqrt{2}$ - the same relation 
as for the total and bare masses in the ``no memory'' state $n=0$! 
The total mass $m_0=m(r_0)$ and the radius $r_0$ in this case are 
related $m_0=4Gr_0$ - twice the horizon size. 

Calculations of Riemann curvature tensor $R^i _{klm}$ and Ricci 
tensor $R_{ik}$ show that if $p<\varepsilon$ ($a\neq b$) there is 
$a$ real singularity at $r=0$. But, surprisingly enough, both 
Riemann and Ricci tensors have finite limits at $r \rightarrow 0$, if 
$\varepsilon =p$ (a=b=1/4G). Therefore we are allowed to introduce 
the so-called topological temperature in the same way as for 
classical black holes. The recipe is the following. One should 
transform the space-time metric by the Wick rotation to the 
Euclidean form and smooth out the canonical singularity by the 
appropriate choice of the period for the imaginary time coordinate. 
The imaginary time coordinate is considered proportional to some 
angle coordinate. In our case the point $r=0$ is already the 
coordinate singularity. The azimuthal angle $\phi$ has the period 
equal to $\pi$. Thus, all other angles should be periodical with 
the period $\pi$. The topological temperature is just the inverse 
of this period. 

The easy exercise shows, that the temperature 
\begin{equation}
T = \frac{1}{2\pi r_0} = \frac{1}{8\pi Gm_0}=T_{BH}
\label{26}
\end{equation}
exactly the same as the Hawking's temperature $T_{BH}$ \cite{5}! 
The very possibility of introducing a temperature provides us 
with the one-parameter family of models with universal distributions 
of energy density and pressure 
\begin{equation}
\varepsilon = p =\frac{1}{16\pi Gr^2},
\label{27}
\end{equation}
the parameter being the total mass $m_0$ or the size $r_0=4Gm_0$.
It should be noted that the two-dimensional part of the metric (\ref{16})
in the limiting case $a=b$ is nothing more but the Rindler's metric,
and the null surface $r=0$ serves as the event horizon. 
The Carter-Penrose diagrams for this case is shown in Fig.11.
\begin{figure}[htbp] 
\vspace*{13pt}
\centerline{\includegraphics{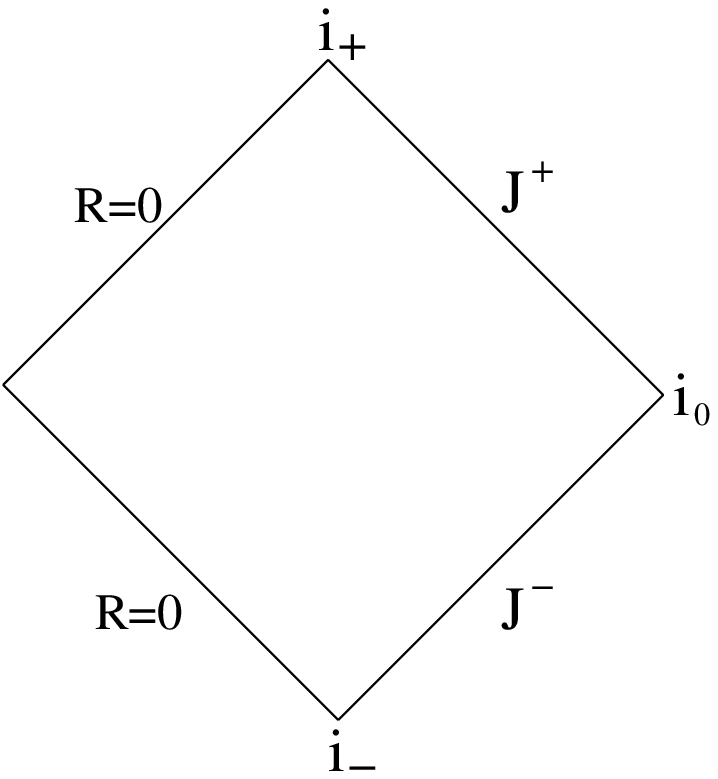}}
\vspace*{13pt}
\caption{}
\end{figure}

We can now develop some thermodynamics for our model. First of all 
we should distinguish between global and local thermodynamic 
quantities. The global quantities are those measured by a 
distant observer. He measures the total mass of the system $m_0$ 
and the black temperature $T_{BH} = T_{\infty}$ and does 
not know anything more. Let us assume that this observer is rather 
educated in order to recognize he is dealing with a black hole 
and to write the main thermodynamic relation 
\begin{equation}
dm = TdS.
\label{28}
\end{equation}
In this way he ascribes to a black hole some amount of entropy, 
namely, the Hawking-Bekenstein value \cite{3,5} 
\begin{equation}
S=\frac{1}{4} \frac{(4\pi r_g)^2}{l^2 _{pl}} =
4\pi Gm_0 ^2 = 
4\pi (\frac{m_0}{M_{pl}})^2
\label{29}
\end{equation}
The local observer who measure distribution of energy, pressure 
and local temperature is also rather educated and writes quite 
a different thermodynamic relation 
\begin{equation}
\varepsilon (r)= T(r)s(r)-p(r)-\mu (r) n(r) 
.
\label{30}
\end{equation}
Here $\varepsilon (r)$ and $p(r)$ are energy density and 
pressure, $T(r)$ is the local temperature distribution, $s(r)$ 
is the entropy density, $\mu (r)$ is the chemical potential, 
and $n(r)$ is the number density of some (quasi)''particles''. 
For the energy density and pressure the local observer 
gets, of course, the relation (\ref{27}), and for the 
temperature - the following distribution 
\begin{equation}
T(r) = \frac{1}{\sqrt{2} \pi r} 
,
\label{31}
\end{equation}
which is compatible with the law $T(r) e^{\frac{\nu}{2}} = const$ 
and the boundary condition $T_{\infty} =T_{BH}$. Such 
a distribution is remarkable in that if some outer layer of our 
perfect fluid would be removed, the inner layers would remain in 
thermodynamic equilibrium. And what about the entropy density? 
Surely, the local observer is unable to measure it directly but 
he can receive some information concerning 
the total entropy from the distant observer. This information and the 
measured temperature distribution (\ref{31}) allows him to 
deduce that 
\begin{equation}
s(r)=\frac{1}{8 \sqrt{2} Gr}
\label{32}
\end{equation}
and
\begin{equation}
s(r)T(r)=\frac{1}{16\pi  Gr^2}
\label{33}
\end{equation}

It is interesting to note that in the main thermodynamic 
equation the contribution from the pressure is compensated 
exactly by the contribution from the temperature and entropy. 
It is noteworthy to remind that the pressure in our classical 
analog model is of quantum mechanical origin as well as the 
black hole temperature. And what is left actually is the 
dust matter we started from in our quantum model, namely, 
\begin{equation}
\varepsilon = \mu n =\frac{1}{16 \pi  Gr^2}
\label{34}
\end{equation}
We may suggest now that the quantum black hole is the 
ensemble of some collective excitations, the black hole 
phonons, and $n(r)$ is just the number density of such 
phonons. 

Knowing equation of state, $\varepsilon = p$, we are able to construct all
the thermodynamical potentials for our system. As an example we show here
how to calculate the energy as a function of the entropy $S$, and the number
particles $N$. By the first law of thermodynamics
\begin{equation}
\label{1lt}
dE = TdS - pdV + \mu dN
\end{equation}
where $T=\left .\frac{\partial E}{\partial S} \right |_{V,N}$ is a 
temperature $p=\left .\frac{\partial E}{\partial V} \right |_{S,N}$ is a
pressure, and $\mu = \left .\frac{\partial E}{\partial N} \right |_{S,V}$ 
is a chemical potential. The energy is additive with respect to the particle
number $N$, hence, $E=Nf(x,y)$ where $x=\frac{S}{N}$ and $y=\frac{N}{V}$.
Since $\varepsilon = \frac{E}{V} = yf(x,y)$ and 
$p = y^2 \frac{\partial f}{\partial y}$ from the equation of state we obtain

\begin{eqnarray}
f = \alpha (x) = n \alpha (x)
\nonumber
\\
\varepsilon = p = n^2 \alpha (x)
\nonumber
\label{f}
\end{eqnarray}

Further,
\begin{eqnarray}
T = n\alpha'(x) 
\nonumber
\\
\mu = n(2 \alpha -x \alpha')
\nonumber
\label{tm}
\end{eqnarray}

But, in any static gravitational field 
$T= T_0/\sqrt{g_{00}}$ and $\mu = \mu_0/ \sqrt{g_{00}}$, so $\mu=\gamma_0 T$,
where $\gamma_0$-some numerical factor. Thus,
\begin{eqnarray}
2 \alpha -x\alpha' = \gamma_0 \alpha',
\nonumber
\\
\alpha(x) = C_0(\gamma_0 + x)^2
\nonumber
\label{x}
\end{eqnarray}
where $C_0$ is a constant of integration. It is easy to see that 
$p/T=1/4C_0$. In our specific model $p/T^2 = \pi/8G$, so $C_0=2G/\pi$.
Moreover, because of the relation $\varepsilon=p=Ts=\mu n$ we know that
free energry $F = E-TS$ is zero. From this we have for the entropy
\begin{equation}
\label{ent}
S = \gamma_0 N
\end{equation}
The black hole entropy equals one fourth of the dimensionless horizon area,
and form this we recover the famous Bekenstein-Mukhanov mass spectrum
\begin{equation}
\label{bm}
m=\sqrt{\frac{\gamma_0}{4\pi}} \sqrt{N}m_{pl}
\end{equation}

We can even calculate the remaining unknown coefficient $\gamma_0$
using the phonon model. Indeed, since the free energy $F=0$ and from the well
know relation
\begin{equation}
\label{free}
F=-T \mbox{ln} \sum_n e^{-\frac{\varepsilon_n}{T}}
\end{equation}
we deduce that the partition function
\begin{equation}
\label{pf}
Z=\sum_n e^{-\frac{\varepsilon_n}{T}} = 1
\end{equation}
Assuming now, that our gravitational phonon are linear oscillators, we have
$\varepsilon_n=\omega(n+\frac{1}{2})$
\begin{equation}
\label{zet}
Z=\frac{e^{-\frac{\omega}{2T}}}{1-e^{-\frac{\omega}{T}}}=1
\end{equation}
and 
\begin{equation}
\label{qu}
e^{-\frac{\omega}{2T}} = \frac{\sqrt{5}-1}{2}
\end{equation}
The energy level is nothing more but bare mass spectrum.
For the black hole mas spectrum we have ($n\gg 1$)
\begin{equation}
\label{mt}
m=\sqrt{2}T n \mbox{ln}\frac{\sqrt{5}+1}{2}=\frac{2n}{8\pi Gm}
\mbox{ln}\frac{\sqrt{5}+1}{2}
\end{equation}
Identifying the level number $n$ with the ``particle'' number $N$ in our
thermodynamical model, we obtain finally
\[
\gamma_0 = \mbox{ln} \frac{\sqrt{5}+1}{2} \approx 0.5
\]
and for the black hole mass spectrum
\begin{equation}
m\approx 0.2 \sqrt{N} m_{pl}
\end{equation}

\section{Conclusion}

And again, the question which already have been asked in Introduction.
Why should we study black holes, for what? I hope, at least something
became clear while reading this paper. The main feature of the black holes
is their universality. It seems it is this that should distinct
quantum black holes from other quantum objects, irrespective of what 
kind of theory the future quantum gravity  will be - fundamental, effective,
induced or some other. In the absence of elaborated quantum theory we have to
construct models. It is desirable that such models were exactly solvable,
because the exact solutions serve as a background of our physical intuition,
which, in turn, allow us to separate the model depending features 
of the object 
under consideration (in our case - quantum black holes) from 
the true fundamental
ones. And, in general, the author believes strongly, that in the future    
the specific 
features of quantum black hole will play the role of some kind of selection 
rule 
for selecting the "correct" quantum theories. It is like the 
"great Fermat theorem"
in mathematics, where numerous attempt to prove it resulted in creation 
of new fields
in the number theory. So,the investigation of quantum black holes is not simply
the exciting intellectual game.

\end{document}